\documentclass[11pt]{iopart}

\usepackage[utf8]{inputenc}

\usepackage[T1]{fontenc}

\usepackage{epsfig}
\usepackage{bm}
\usepackage{color}
\usepackage{verbatim}

\usepackage{iopams}
\expandafter\let\csname equation*\endcsname\relax
\expandafter\let\csname endequation*\endcsname\relax
\usepackage{amsmath}
\usepackage{amssymb}

\usepackage{bbm}

\usepackage{graphicx}
\usepackage{fancybox}

\graphicspath{{./graphics/}}

\newcommand{\op}[1]{\ensuremath{\Hat{#1}}}

\newcommand{\bra}[1]{\ensuremath{\langle{#1}\vert}}
\newcommand{\ket}[1]{\ensuremath{\vert{#1}\rangle}}

\newcommand{\abs}[1]{\ensuremath{\mathinner{\lvert#1\rvert}}}

\renewcommand{\vec}[1]{\ensuremath{\boldsymbol{#1}}}
\newcommand{\expect}[1]{\ensuremath{\langle{#1}\rangle}}

\renewcommand{\b}{\ensuremath{\beta}}
\renewcommand{\d}{\ensuremath{\delta}}
\newcommand{\D}{\ensuremath{\Delta}}
\newcommand{\dd}{\ensuremath{\text{d}}}
\renewcommand{\e}{\ensuremath{\epsilon}}
\newcommand{\E}{\ensuremath{\boldsymbol{e}}}
\newcommand{\g}{\ensuremath{\gamma}}
\newcommand{\G}{\ensuremath{\Gamma}}

\renewcommand{\P}{\ensuremath{\text{P}}}
\newcommand{\s}{\ensuremath{\sigma}}
\newcommand{\w}{\ensuremath{\omega}}

\newcommand{\dt}{\ensuremath{\delta_t}}
\newcommand{\Dt}{\ensuremath{\Delta_t}}

\newcommand{\TauC}{\ensuremath{\tau_{c}}}

\newcommand{\grass}[1]{\ensuremath{\mathfrak{#1}}}
\newcommand{\hrass}[1]{\ensuremath{\overline{\mathfrak{#1}}}}

\newcommand{\boxx}[3]{
\cornersize{0.15}
	\Ovalbox{\ensuremath{#1}}_{\ensuremath{\, #2}}^{\ensuremath{\, #3}}
}
\newcommand{\dboxx}[2]{
\cornersize{0.15}
      \doublebox{\ensuremath{#1_{\, #2}}}
}
\usepackage{xcolor}

\definecolor{plot_a}{rgb}{0.2472, 0.2400, 0.6000}
\definecolor{plot_b}{rgb}{0.6000, 0.2400, 0.4429}
\definecolor{plot_c}{rgb}{0.6000, 0.5470, 0.2400}
\definecolor{plot_d}{rgb}{0.2400, 0.6000, 0.3369}

\newcommand{\Grass}{\ensuremath{\Psi}}
\newcommand{\Hrass}{\ensuremath{\overline{\Psi}}}
\newcommand{\IdOp}{\ensuremath{\boldsymbol{\mathbbm{1}}}}
\newcommand{\HamEff}[1]{\ensuremath{H_{\text{#1}}^{\zeta}}}
\newcommand{\normal}[1]{\ensuremath{\,\mathord{:} #1 \mathord{:}\,}}
\newcommand{\be}{ \begin{equation} }
\newcommand{\ee}{ \end{equation} }
\newcommand{\bea}{ \begin{eqnarray} }
\newcommand{\eea}{ \end{eqnarray} }

\begin{document}

\title{Nonequilibrium quantum dynamics of the magnetic Anderson model}

\author{D Becker$^{1,2}$, S Weiss$^{1,3}$, M Thorwart$^{1}$ and D
Pfannkuche$^{1}$}

\address{$^1$I. Institut f\"ur Theoretische Physik, Universit\"at Hamburg,
Jungiusstra{\ss}e 9, 20355 Hamburg, Germany}
\address{$^2$ Departement Physik, Universit\"at Basel, Klingelbergstrasse 82,
CH-4056 Basel, Switzerland}
\address{$^3$ Institut f\"ur Theoretische Physik IV, Heinrich Heine Universit\"at
D\"usseldorf, Universit\"atsstr.\ 1, 40225 D\"usseldorf, Germany
}

\ead{dbecker@physnet.uni-hamburg.de}

\begin{abstract}
We study the nonequilibrium dynamics of a spinful single-orbital quantum dot
with an incorporated quantum mechanical spin-1/2 magnetic impurity. Due to the
spin degeneracy, double occupancy is allowed and Coulomb interaction together
with the exchange coupling of the magnetic impurity influence the dynamics.
By extending the iterative summation of real time path integrals (ISPI)
to this coupled system, we monitor the time-dependent nonequilibrium current and
the impurity spin polarization to determine features of
the time-dependent nonequilibrium dynamics. We especially focus on the deep
quantum regime, where all time- and energy scales are of the same order of
magnitude and no small parameter is available. We observe a significant
influence of the 
nonequilibrium decay of the impurity spin polarisation both in presence and
in absence of Coulomb interaction. The exponential relaxation is faster
for larger bias voltages, electron impurity interactions and temperatures. We
show that the exact relaxation rate deviates from the corresponding
perturbative result. In addition, we study in detail the 
impurity's back action on the charge current and find a reduction of the
stationary current for increasing coupling to the impurity. Moreover, 
our approach allows to systematically distinguish
mean-field Coulomb and impurity effects from the influence of quantum
fluctuations and flip-flop scattering, respectively. In fact, we find a
local maximum of the current for a finite Coulomb
interaction due to the presence of the impurity.

\end{abstract}

\submitto{\NJP}

\maketitle

\section{Introduction}

Diluted magnetic semiconductors \cite{furdyna:R29, 0268-1242-17-4-310,
RevModPhys.78.809, Beaulac2008} are an important class of materials for the
spintronics community since they combine properties of (ferro-)magnets and
semiconductors.  Thus they allow for the all-electrical control of the magnetic
degrees of freedom of spintronic devices.  The magnetisation of semiconducting
devices is mainly caused by the interaction of magnetic impurities in the sample
with itinerant charge carriers. To understand the microscopic details of these
magnetic materials, reduced model systems are required in which the individual
constituents are well under control. This allows to study fundamental questions
concerning the interplay of coherent quantum dynamics and dissipation under
nonequilibrium conditions. An ideal candidate for such a model system is a small
quantum dot which connects metallic leads and also carries a magnetic degree of
freedom described by a quantum mechanical spin ({\em magnetic Anderson model}). 

Magnetic quantum dots have been studied experimentally in ensembles which are
particularly suited for the investigation by laser and electromagnetic fields
\cite{mackowski:3337, PhysRevB.71.035338, 0295-5075-81-3-37005,
PhysRevLett.102.177403, Zutic2009, PhysRevB.81.245315, Ochsenbein2009}. They are
designed with standard lithographic methods and are technologically well
established. Moreover, embedding individual Mn ions into quantum dots and
studying the electrical properties is possible. \cite{PhysRevLett.97.107401,
PhysRevLett.98.106805, Hanson2008}. Small quantum dots with few charge carriers
and a single magnetic impurity may become important candidates for efficient
high density spintronic devices.
Although Mn ions in quantum dots have usually spin larger than $S =
1/2$, there are good reasons to study the low spin case first. From
a methodological point of view, this simple model is an ideal basis to
develop the numerical tools necessary to treat the real-time dynamics 
of more general coupled electron-impurity spin systems as well. 
In addition, the magnetic Anderson model serves as a generic model to study electronic 
transport through magnetic atoms or molecules placed between the tip of a
scanning tunnelling microscope and a substrate
\cite{Wiebe07,Khajetoorians10,Khajetoorians11}. Likewise, our model is a
phenomenological basis for microscopic studies of molecular junctions  based on
organic radicals \cite{Herrmann}. We mention the switching of the spin
state of the central iron ion in a single molecular complex in a double layer
on gold by a low temperature scanning tunnelling microscope \cite{Berndt} for
which our model also is applicable. Finally, it also mimics 
features of the dynamics of electrons in quantum dots that are subject to
hyperfine interactions with the nuclei of the host material.
In that respect, the understanding of the electron-impurity coupling and its
influence on the dephasing and relaxation times of qubits is essential for the
experiments realized in Refs.\  \cite{Hanson2007, Churchill2009}. 

Nonequilibrium quantum transport in strongly correlated systems continues to
remain a challenging problem. Especially, reliable theoretical treatments prove
to be difficult  when no small parameter is present in the system, i.e., in the
deep quantum regime. Approximate methods are often based on advanced
perturbative treatments such as quantum master equations
\cite{kouwenhoven_ganz:1996,PhysRevB.77.195416,PhysRevB.50.18436}, which base on
Markovian approximations and weak tunnelling coupling. The renormalisation group
(RG) approach \cite{PhysRevLett.84.3686, PhysRevLett.90.076804,
PhysRevB.69.155330, PhysRevLett.95.056602, PhysRevB.77.125309,10.1140Schoeller}
as well as the functional RG method are able to capture essential nonequilibrium
features \cite{jakobs:150603, PhysRevB.75.045324,
0295-5075-90-3-30003,PhysRevB.81.195109}. Real time density matrix RG's
\cite{PhysRevB.79.235336, 1742-5468-2004-04-P04005, PhysRevLett.93.076401,
PhysRevB.70.121302} require to represent the system plus the macroscopic
reservoirs by a large but finite lattice. Due to the possible appearance of
finite size effects, the maximal propagation times are limited.

Quantum Monte-Carlo (QMC) concepts are a priori numerically exact and have been
adopted to nonequilibrium quantum transport
\cite{muehlbacher:176403,PhysRevB.78.235110, werner:035320,
PhysRevB.81.035108,schiro:153302}. As opposed to the fermionic sign problem,
which shows up in equilibrium simulations of quantum many-body systems, the
real-time (nonequilibrium) QMC weight function is itself highly oscillatory and
causes the dynamical sign problem. Reaching the stationary nonequilibrium state
at asymptotic times remains notoriously difficult \cite{LichtensteinRMP,
muehlbacher:176403, PhysRevB.78.235110, werner:035320,
PhysRevB.81.035108,schiro:153302}. An analytic continuation to the ``complex
voltage plane'' via imaginary ``Matsubara voltages'' attempts to circumvent the
dynamical sign problem. However, the postprocessed back transformation to real
times is plagued by numerical instabilities
\cite{PhysRevLett.99.236808,EckelNJP2010}.
Recently, long-time and steady state results have been obtained by a QMC
sampling of the diagrammatic corrections to the
non-crossing approximation \cite{GullPRB2011}, stating the reduction of the sign problem.

In this work, we adopt the method of the iterative summation of path integrals
(ISPI), see Ref.~\cite{weiss:195316} to the magnetic Anderson model. This
approach evaluates a real time path integral expression for the Keldysh
partition function in a numerically exact manner and is particularly suitable
for nonlinear transport. Recently, this scheme has been carefully verified by a
comparison to existing approximations in the appropriate parameter regimes for
the single-impurity Anderson model \cite{weiss:195316}.  Furthermore, the
prediction of a sustained Franck-Condon blockade in the deep quantum regime has
been reported based on ISPI data as well \cite{huetzen2012}. In contrast to the
stochastic evaluation of the real time path integral in the QMC approach, the
ISPI scheme calculates the real time path-integral deterministically. Hence the
sign problem is avoided. The fact that metallic leads at finite temperature
suppress all electronic correlations exponentially beyond a finite memory time
window is exploited by the ISPI method. This allows for an iterative propagation
of an augmented reduced density operator of the system to arbitrary long times.
By construction, the technique is limited to finite temperatures and/or finite
bias voltages and not too large system sizes. Whenever numerical results are
converged with respect to the memory time, they are numerically exact. Recently,
Segal et al. \cite{PhysRevB.82.205323} have provided an alternative formulation
of this approach in terms of Feynman-Vernon-like influence functionals. 

We theoretically investigate the real time dynamics of a single-level quantum
dot containing a magnetic impurity with spin-1/2. Exchange interaction with
on-dot electrons results in dissipation, induced by the metallic leads to the
localised impurity in the quantum dot. The deep quantum regime, characterised by
the absence of a small parameter is in the focus of the present article, i.e.,
all interaction strengths are of the same order of magnitude. In particular, we
are interested in the nonlinear transport regime, where a finite bias voltage is
applied and linear response theory is no longer applicable. The real time
relaxation of the impurity spin as a function of various system parameters is
investigated. Due to the additional degree of freedom of the magnetic impurity,
an extension of the ISPI scheme  \cite{weiss:195316} is necessary.  Although the
inclusion of a magnetic impurity affects the single particle dynamics in the
first place, coupling of the impurity to the electronic density on the dot
renders this extension nontrivial. Our accurate results show that in the
considered cross-over regime, the nonequilibrium charge current significantly
influences the quantum relaxation dynamics of the impurity. Likewise, the back
action of the impurity dynamics on the nonequilibrium current becomes
significant. Most importantly, we clearly show that this crossover regime is not
accessible by perturbative means.

The structure of the article is as follows. After introducing the model in
Sec.~\ref{sec:Model}, we express the Keldysh partition function as a real time
path integral in Sec.~\ref{sec:ISPI} and show how to evaluate this path sum by a
deterministic iteration scheme. We calculate in Sec.~\ref{sec:results} the
stationary transport current and the impurity spin dynamics. Some of the results
are compared to outcomes of perturbative methods for the appropriate parameter
combinations. We analyse the influence of the nonequilibrium current on the
impurity relaxation.  Furthermore, we present results of the stationary
tunnelling current in the deep quantum regime, where rate equation results are
not reliable. The dependence of the relaxation rate on various model parameters
is presented. 

\section{Model system} \label{sec:Model}

\begin{figure}
\begin{center}
\includegraphics{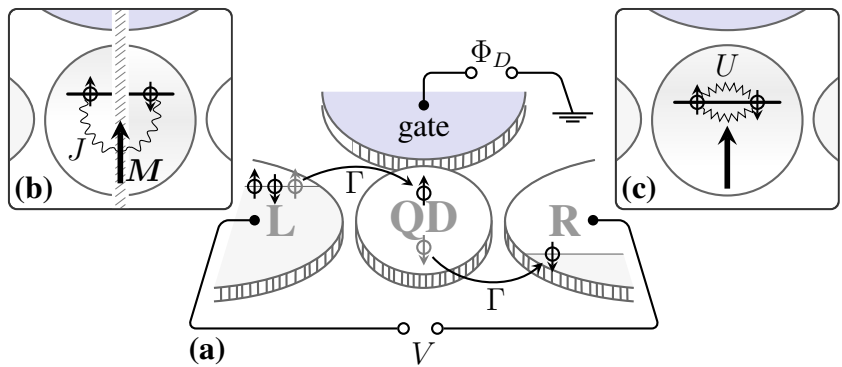}
\caption{\label{fig:model}
(a) Two metallic leads are coupled to a quantum dot (QD)
via tunnelling barriers. The gate electrode allows to tune the electrostatic
potential $\Phi_D$ and the bias voltage $V$ induces an electron current. The
QD has a single orbital electronic level which is empty or
occupied either by one electron in spin up or down
 state (b) or by two electrons in a Coulomb-interacting
singlet state (c). Moreover, the QD carries 
a localised magnetic impurity $\vec{M}$ with spin 1/2. In (b), a 
{\em single\/} dot electron interacts with $\vec{M}$ via an exchange 
interaction $J$, while in (c) only Coulomb interaction is present for
double occupation.}
\end{center}
\end{figure}

We extend the single-impurity Anderson model for the electronic degree of
freedom of a quantum dot (QD) coupled  to two metallic leads (L/R) via
tunnelling barriers in order to study magnetic quantum dots, see Fig.\
\ref{fig:model} for a sketch. We assume equal tunnelling barriers at the left and
right side, the generalisation to the asymmetric case is straightforward. The
total Hamiltonian is given by $H = H_{\text{dot}} + H_{\text{leads}} + H_{T}$
where we use units such that $\hbar=1$. The Hamiltonian of the magnetic dot 
\begin{equation}
 H_{\text{dot}} = H_{\text{dot}}^{\text{el}} + 
H_{\text{imp}} + H_{\text{int}} 
\end{equation}
includes the electronic part $H_{\text{dot}}^{\text{el}}$ and the part
$H_{\text{imp}}$ to describe the spatially fixed magnetic impurity as well as
the coupling term between electron and impurity denoted as $H_{\text{int}}$. 

We write the electronic part of the QD as 
\be
H_{\text{dot}}^{\text{el}} = H_{\text{dot}}^{0} +
H_{\text{dot}}^{U}= \sum_{\s} \e_{\s}
d_{\s}^\dagger d_{\s}+ U d^\dagger_{\uparrow}
d_{\uparrow} d^\dagger_{\downarrow} d_{\downarrow} \, ,
\label{Hdotel}
\ee
where the operator $d_{\s}$ annihilates a dot electron with spin $\s$. The dot
is formed by a single spin degenerate level, 
which can be controlled by a gate electrode that shifts the electrostatic
potential of the dot $\Phi_D$.
Hence, the electronic degree of freedom can assume four values $\{ 0, \uparrow,
\downarrow, \text{d} \}$, indicating whether the
dot is empty ($0$), contains one electron with spin $\s =
\uparrow,\downarrow = \pm 1$ and energy $\e_{\s} = \Phi_D + \s \D / 2$, or is in
a spin singlet state with double occupation (d). The Coulomb repulsion is
modelled via $U > 0$ 
when the dot is in the doubly occupied state d. The Zeeman level splitting $\D$
might be present due to 
possible external or internal crystallographic magnetic fields. 

The magnetic impurity (quantum spin) is included via the Hamiltonian 
\be
H_{\text{imp}} = \frac{\D_{\text{imp}}}{2} \tau_z \, ,
\ee
 with the Zeeman energy $\D_{\text{imp}}$. Spin operators of the impurity are
given in terms of the Pauli matrices $\tau_{x,y,z}$
with $\tau_{\pm} = \tau_{x} \pm i \tau_{y}$. The impurity spin $\tau$ and the
dot electron spins $\sigma$ are
coupled by the exchange interaction of strength $J$ which is 
captured by the Hamiltonian 
\begin{equation}
\label{eqn:model:H:int}
H_{\text{int}} = 
\underbrace{J \tau_z  (d^\dagger_{\uparrow}
d_{\uparrow} - d^\dagger_{\downarrow} d_{\downarrow})}_{\displaystyle
H_{\text{int}}^{\parallel}} + 
\underbrace{\frac{J}{2} (\tau_{+} d^\dagger_{\downarrow} d_{\uparrow}  +
\tau_{-} d^\dagger_{\uparrow} d_{\downarrow})}_{\displaystyle
H_{\text{int}}^{\perp}} \, .
\end{equation}
While the first term shifts the single-particle
energies, the second term induces mutual flips of an electron- and the impurity
spin, which we denote as {\em flip-flop processes\/}. The interaction vanishes for 
double occupation of the dot, then  
two electrons form a spin-singlet state.

As usual, we model the noninteracting leads by $H_{\text{leads}} = \sum_{\vec{k} \s p}
\e_{\vec{k}} c^\dagger_{\vec{k} \s p} c_{\vec{k} \s p}$, 
where $c_{\vec{k} \s p}$ annihilates an
electron with spin $\s$ and wave vector $\vec{k}$ in lead $p = \text{L},
\text{R} = \pm 1$. A bias voltage $V$ is symmetrically applied between the
(thermal) leads and shifts their electrochemical potentials 
such that $\mu_{p} = p e V / 2$. Finally, the tunnelling Hamiltonian
$H_T = \sum_{\vec{k} \s p} \g d^\dagger_{\s} c_{\vec{k} \s p} + \g^*
c^\dagger_{\vec{k} \s p} d_{\s}$ describes the energy- and spin-independent tunnelling of
electrons with the amplitude $\g$. We assume a constant density of
 states $\varrho(\e_{\text{F}})$
around the Fermi energy  and work in the wide-band
limit. Then, the tunnelling is parametrised by
the parameter $\G := 2 \pi \abs{\g}^2 \varrho(\e_{\text{F}})$. 

\section{The ISPI approach} \label{sec:ISPI}

In order to determine the nonequilibrium electron current and to 
study the relaxation properties of the impurity, we generalise 
 the approach of the \emph{iterative summation of  path integrals} (ISPI)
of Ref.\ \cite{weiss:195316}. This is a nontrivial step and requires to
incorporate the impurity quantum dynamics consistently within the quantum 
many-body formalism.

\subsection{Path integral, generating function and short-time propagator}

The ISPI approach is based on the Keldysh partition function  $\mathcal{Z} =
\Tr\{ U_K [H] \rho (-\infty) \} = 1$, where $U_K [H] = T_{K} \exp
\{-i \int_{\mathcal{K}} H dt \}$ is the propagator along the Keldysh contour
$\mathcal{K}$.  The time ordering operator is $T_K$ and $\rho
(-\infty)$ is the
density matrix of the system's initial state \cite{Keldysh64,
kamenev-2009-58}. A generating function 
$\mathcal{Z}[\eta (t)]$ is constructed to calculate the expectation
value of an Operator $O(t)$ via 
\begin{equation}\label{eqn:FunctionalDerivative}
\expect{O}_t = \left.\frac{\d \ln \mathcal{Z} [\eta(t)]}{\d \eta
(t)}\right\vert_{\eta = 0}
\end{equation}
with $\mathcal{Z} [\eta(t)]= \Tr\{ U_K [ H + i 
\eta(t) O ]\, \rho(-\infty)\}$. Then, $\mathcal{Z}[\eta (t)]$ is represented by 
a (fermionic) path integral. Throughout this work, we assume an initially $(t_i =-\infty)$ empty 
quantum dot prepared in the spin-up impurity state. The full Keldysh time interval $\Dt := t_f - 
t_i = N \dt$ is discretised into time steps $\dt$. A short-time
propagator $U_{\dt}$ is introduced such that it is related to the exact propagator
$U(t+\dt, t) = U_{\dt} + \mathcal{O} (\dt^2)$. Subsequently, 
 a complete basis of fermionic coherent states is inserted between every two
short-time propagators, accounting for the Coulomb 
and the flip-flop interactions.

To construct a particular $U_{\dt}$, we separate the total Hamiltonian $H$ into
the diagonal part $H_0$ and a nondiagonal part $H_1$ with respect to appearing
dot operators $d_\sigma$ and $d_\sigma^\dagger$. This yields $H=H_0+H_1$ with  
\begin{eqnarray}
H_0 & = & H_{\text{dot}}^{0} + H_{\text{imp}} + H_{\text{int}}^{\parallel}
+H_{\text{leads}} \, , \nonumber \\
\label{eqn:model:H:ND}
H_1 &= &H_{\text{dot}}^{U} + H_{\text{int}}^{\perp} +H_T \, .
\end{eqnarray}
In the interaction picture, the full real-time propagator from the initial to 
the final time can be written as 
\bea
U(t_f , t_i) &=&\sum_{N=2}^{\infty} \left( -i \right)^{N-2}
\int_{t_i}^{t_f} \dots \int_{t_{N-2}}^{t_f} dt_2 dt_3 \dots dt_{N-1} \nonumber
\\
& & \times U_0(t_f, t_{N-1}) H_1 \dots H_1 U_0 (t_3, t_2) H_1 U_0(t_2, t_1) \, ,
\eea
where the Keldysh contour is divided into $N-1$ pieces of free
propagation by $U_0 (t_{k+1}, t_k)$ that are connected by $N - 2$
interaction vertices $-i H_1 dt_k$ acting during the 
transition time $dt_k$. Here, we define $t_1 := t_i$ and $t_{N} :=
t_{f}$. When $t_{k+1}=t_k + \dt$ and in the limit of very small $\dt$, the
system can either propagate freely  during
$\dt$ via the free propagator $U_0 (\dt) = U_0(t_{k+1}, t_k) = 
U_0(t_{k+1}- t_k)$ or undergo at most one transition induced by $H_1$ within the
interval $0
< t' < \dt$. Hence, the short time propagator takes the form 
\begin{equation}\label{eqn:ShortTimePropInt} 
U_{\dt} = U_0 (\dt) - i \int_0^{\dt}\!\!\!\!\! dt'
U_0 (\dt - t') H_1 U_0 (t') =\normal{U_0 (\dt) \left(\IdOp - i H_1 \dt \right)}
+ \mathcal{O} (\dt^{2}) \, ,
\end{equation}
where the commutator  $[U_0 (t), H_1] =  \mathcal{O} (t)$ is
used to bring $H_1$ to the right of $U_0 (t)$. The 
remaining time integral is evaluated exactly up to order $\mathcal{O}(\dt^{2})$. A
comparable error is obtained by normal ordering, denoted by colons in Eq.~\eqref{eqn:ShortTimePropInt}. 

\subsection{Discrete Hubbard-Stratonovich Transformation} \label{subsec:HSTrafo}

Next, we treat the Coulomb term $H_{\text{dot}}^{U}$ in a 
similar way as in Ref.\
\cite{weiss:195316}. Since $[H_0,H_{\text{dot}}^{U}]=0$, 
the propagator for the system with Coulomb interaction factorises 
into a free part $U_{0} (\delta_t)$ and the interacting part
$\exp\{ -i H_{\text{dot}}^{U} \dt\}$. We apply the
Hubbard-Stratonovich (HS) transformation and obtain 
\be\label{eqn:HSTransformation}
\exp \{ -i  H_{\text{dot}}^U \dt \}=\frac{1}{2}\sum_{\zeta =
\pm 1}\exp \left\{-i  \left[\frac{U}{2}(n_{\uparrow} + n_{\downarrow})
+i  \zeta\! \lambda_{\dt} \, (n_{\uparrow} - n_{\downarrow}) \right] \dt
\right\} 
\ee
with the HS parameter $\lambda_{\dt}$ defined via 
\be \label{eqn:HSTransformation:b}
\lambda_{\dt} \dt = \sinh^{-1}  \sqrt{\sin[U \dt/2]}  + i\sin^{-1}  \sqrt{\sin[U
\dt/2]} \, .
\ee
The variable $\zeta=\pm 1$ is interpreted as a fluctuating Ising-like spin
field. Note that
the solution given in Eq.~\eqref{eqn:HSTransformation:b} is uniquely defined  as
long as $0 \le U \dt \le \pi$. By this step, 
the interacting problem with local-in-time Coulomb repulsions is effectively
mapped to 
a noninteracting problem, at the price that the appearing spin fields interact
over a finite time range with each other. In general, the condition 
$\delta_t\Gamma\ll 1$ is needed to minimise the systematic error of order
$\delta_t^2$. If $\delta_t$ is bounded from below, however,
$U$ should be adjusted in agreement with the condition in
Eq.~\eqref{eqn:HSTransformation:b}. The exponential  in Eq.\ 
\eqref{eqn:HSTransformation} commutes with $H_0$ and we can write
the full short-time propagator as $\exp\{-i (H_0 +
H_{\text{dot}}^{U})\dt\} = 1/2 \sum_{\zeta=\pm 1} \exp\{ -i\, \HamEff{0}
\dt\}$, thereby absorbing the classical part of the Coulomb
interaction into the single-particle dot
energies according to 
\begin{equation}\label{eqn:EffectiveDotEnergy}
\e_{\s}^{\zeta}(\dt) = \e_{\s} + \frac{U}{2} + i  \sigma \zeta
\lambda_{\dt} \, .
\end{equation}
In passing, we note that due to the imaginary energy
component, $\HamEff{0}$ should not be considered as a Hamiltonian. Instead, we
obtain a short-time propagator by enforcing normal ordering, i.e.,  
\begin{equation}\label{eqn:ShortTimeCoulomb}
U_{\dt}^{U}:=\frac{1}{2}\sum_{\zeta = \pm 1}\normal
{\exp\{-i H_0^{\zeta} \dt\}} \, .
\end{equation}
Combining the short-time propagators again into the full path integral, 
the path sum extends over all tuples $\{\zeta\}
\!:=\! (\zeta_N,\ldots,\zeta_1)$ of the HS fields.

\subsection{The remaining interaction terms} \label{subsec:AddRemainInt}

The tunnelling term $H_{T}$ is quadratic in the number of fermions but
contains both dot and lead operators. Therefore, the stationary state of the
isolated system is in general not an eigenstate of the system with tunnelling. 
However, for arbitrary electronic coherent and impurity states
$\ket{\Psi^{\tau}} \equiv \ket{\Psi} \ket{\tau}$, the matrix
elements can be written as 
\be\label{eqn:TwoPropMElements}
\bra{\Psi^{\tau'}}
\normal{\exp\{-i H_0^{\zeta} \dt\}\left(\IdOp-i H_T
\dt\right)}\ket{\Psi^{\tau}} = \bra{\Psi^{\tau'}} \normal{\exp\{-i
(H_0^{\zeta} + H_T) \dt\}}
\ket{\Psi^{\tau}} + {\cal O} (\dt^2) \, .
\ee
Hence, 
the short time propagator is obtained by adding the
coupling term $H_{T}$ to $\HamEff{0}$ in Eq.\ 
\eqref{eqn:ShortTimeCoulomb}. 
To be specific, we introduce the total electronic coherent state as 
\begin{equation}\label{eqn:}
\ket{\Psi} 
\equiv
\prod_{\vec{k}\s p} (1 - \grass{c}_{\vec{k}\s p} c^\dagger_{\vec{k}\s p})
	\prod_{\s} (1 - \grass{d}_{\s} d^\dagger_{\s}) \ket{0}\, ,
\end{equation}
where $\grass{d}_{\s}$ and $\grass{c}_{\vec{k} \s p}$ are Grassmann  numbers for
dot and lead electrons.

Flip-flops of the electron- and impurity spin are mediated by the non-diagonal
exchange coupling term $H_{\text{int}}^{\perp}$ in Eq.~\eqref{eqn:model:H:ND} . 
According to Eq.~\eqref{eqn:ShortTimePropInt} it enters the short time
propagator as 
\begin{equation}\label{eqn:ShortPropFinal}
U_{\dt} = \frac{1}{2} \sum_{\zeta=\pm 1} \normal{\exp\{-i (H_0^{\zeta} +
H_T)\dt\}\left(\IdOp -i H_{\text{int}}^{\perp} \dt \right)} \, .
\end{equation}
In fact, this structure of the short-time propagator motivates our choice of a
``mixed'' basis of electron coherent states $\ket{\Psi}$ and impurity states 
$\ket{\tau}$. Most importantly,  a straightforward derivation of a path integral
in this basis becomes possible as the paths are separated into parts that
contribute to the matrix element $\bra{\Psi^{\tau'}} U_{\dt} \ket{\Psi^{\tau}}$
either for aligned ($\propto \IdOp$) or opposite ($\propto
H_{\text{int}}^{\perp}$) impurity spin orientations. This form is particularly
useful with respect to a numerical summation over discrete impurity paths. An 
equivalent short-time propagator in form of a single exponential
$\normal{\exp\{ - i H^{\zeta} \dt \}}$ with $H^{\zeta} := \HamEff{0} +
H_{T} + H_{\text{int}}^{\perp}$ is much more cumbersome. 

In contrast, the short-time propagator $\IdOp - i H^{\zeta} \dt$, though exact 
up to ${\cal O}(\dt^2)$, is not convenient to construct a path integral.
Consider, the case of opposite impurity spin states $\tau \ne \tau'$. The phase
accumulated by the free propagation between two instantaneous flip-flop events
is missing, i.e.,
\begin{equation}\label{eqn:WrongFlipFlopResult}
		\bra{\Psi^{\tau'}} \IdOp - i H^{\zeta} \dt \ket{\Psi^{\tau}}_{\tau \ne \tau'}
	=	-\frac{i J \dt}{2} (\d_{\tau',\tau+1} \hrass{d}_{\downarrow}'
		\grass{d}_{\uparrow} +\d_{\tau', \tau - 1} \hrass{d}_{\uparrow}'
		\grass{d}_{\downarrow})\E^{\Psi'\Psi}\, .
\end{equation}
Such a propagator would only be correct if the transition process lasted the 
whole time span $\dt$ instead of being instantaneous. In the resulting path
integral, the system does not propagate freely between consecutive flip-flop
processes with the resulting continuous limit $\delta_t\to 0$ being unphysical. 

\subsection{Constructing the full path integral} \label{subsec:fullpi}

The path integral for the generating function $\mathcal{Z}[\eta]$ is obtained by using 
Eq.\ \eqref{eqn:ShortPropFinal} and the corresponding Grassmann fields $\Psi$ 
as well as the discrete paths $\{\tau\}$ and $\{\zeta\}$. This yields 
\begin{equation}\label{eqn:KeldyshPathIntZ0}
\mathcal{Z}[\eta]=\sum_{\{\tau, \zeta\}}
\int\!\! \mathcal{D} [\Hrass \Grass]\, (-1)^{\ell} \left(-\frac{i J
\dt}{2}\right)^{m} P [\{\tau\}] e^{i S [\{\Hrass, \Grass, \tau, \zeta\}]} \, .
\end{equation}
where the path sums over impurity and HS 
spin-fields are performed over the $2N$-tuples $\{\tau_j\} = (\tau_{2 N}, \ldots,
\tau_{1})$ and $\{\zeta_j\} = (\zeta_{2 N}, \ldots, \zeta_{1})$ with 
$\tau_j, \zeta_j=\pm 1$.
Within an impurity path $\{\tau\}$, $m$ flip-flop transitions occur on the 
Keldysh contour, where $\ell$ of them lie on the lower branch. 
We discuss the building blocks of Eq.~\eqref{eqn:KeldyshPathIntZ0} in the following.

\begin{figure}
\begin{center}
\includegraphics{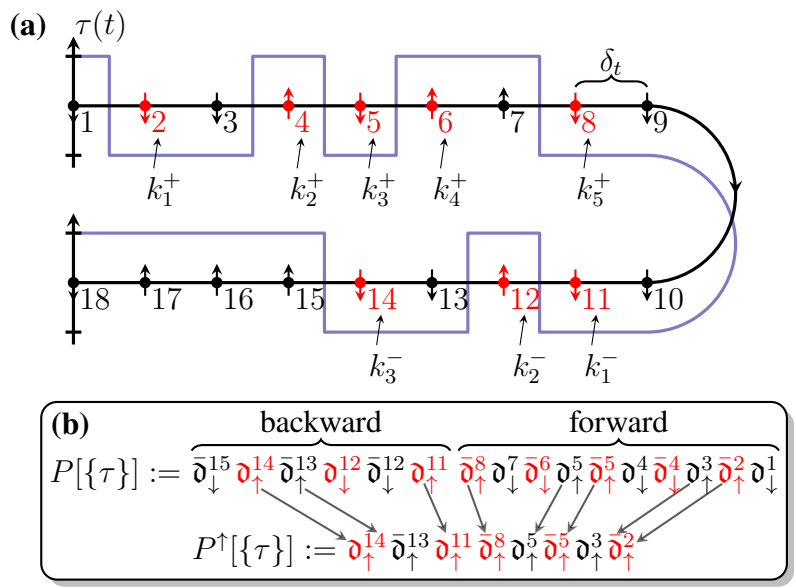}
\caption{
\label{fig:ConstructP}
(a) Exemplary impurity path (blue line) for which the flip-flop polynomial is
constructed. The
Keldysh contour is divided into $N - 1 = 8$ segments of length $\dt$ between $2
N = 18$ time
vertices. The impurity path (tuple of black and red arrows) realizes $m=8$
flip-flops with five flip-flops
on the forward and $\ell = 3$ flip-flops on the backward branch. First, the flip
index tuples
$T_{\text{flip}}^{\pm}$ are constructed by assigning to each flip-flop the
index of the Keldysh time that \emph{is later with respect to the real time}.
Hence, if
two consecutive spins have opposite orientations, the corresponding flip-flop
 has the time index of the spin on the right-hand side of the flip
(marked in red).
We have $T_{\text{flip}}^+ = (8, 6, 5, 4, 2)$ and $T_{\text{flip}}^-
= (14, 12, 11)$. The polynomial written in (b) follows upon using 
Eq.\ \eqref{eqn:KeldyshFlipFlopPolynomial}. We note that the electron's spin
flips in the opposite
direction as compared to the impurity (not shown).}
\end{center}
\end{figure}

In Eq.~\eqref{eqn:KeldyshPathIntZ0}, the action $S = S_{\text{imp}} +
S_{\text{el}}
+ S_O [\eta]$ is given by the sum of the free impurity action and the electronic
action. 
Due to possible source terms, depending on the observable of interest $O(t)$,
$S_O [\eta]$ is added when necessary (see below). The electronic action contains
all 
coupling terms of the electrons to the impurity, to the leads (via the
tunnelling) and to the HS fields. In particular, we have the action of the free
impurity 
\begin{equation}\label{eqn:ContinuousAction:imp}
S_{\text{imp}}
=-\frac{\Delta_{\text{imp}} \dt}{2} \sum_{k = 2}^N ( \tau_k  -
\tau_{2 N - k + 1} )
=-\frac{\Delta_{\text{imp}}}{2}
\int_{\mathcal{K}}dt\, \tau (t)\, ,
\end{equation}
with the discrete and continuous representation given as the first and the
second expression, respectively.
This equation also illustrates how a well-defined continuous limit of a discrete
path can be obtained (cf.\ Fig. \ref{fig:ConstructP}). The electronic action 
$S_{\text{el}}$ can be represented as
\begin{equation} \label{eqn:ContinuousAction:EL}
S_{\text{el}}=S_{\text{dot}}^{\text{el}}+S_{\text{leads}}+S_T
=\int_{\mathcal{K}} dt dt'\,
\Hrass(t)  (G^{\text{el}})^{-1}_{t,t'} \Grass(t')
\end{equation}
with the inverse electronic Keldysh Green's function
$(G^{\text{el}})^{-1}_{t,t'}$ naturally given in terms of the action
$S_{\text{el}}$ with the three contributions 
\begin{eqnarray}
\label{eqn:ContinuousAction:dot}
S_{\text{dot}}^{\text{el}}&=&\sum_{\s}\int_{\mathcal{K}} dt
\hrass{d}_{\s} (t) [i \partial_t -E_{\s} (t)]\grass{d}_{\s} (t)\, , \nonumber \\
S_{\text{leads}} &=&\sum_{\vec{k} \s p}\int_{\mathcal{K}} dt 
\hrass{c}_{\vec{k} \s p} (t)[i \partial_t - \e_{\vec{k}}]\grass{c}_{\vec{k} \s
p} (t)\, , \nonumber
\\
S_T&=&\sum_{\vec{k} \s p}\int_{\mathcal{K}} dt [\g\hrass{d}_{\s}
(t)\grass{c}_{\vec{k} \s p} (t)
+\g^* \hrass{c}_{\vec{k} \s p} (t) \grass{d}_{\s} (t)] \, .
\end{eqnarray}
We note that one of the time integrations in Eq.\
\eqref{eqn:ContinuousAction:EL} is trivial, since $(G^{\text{el}})^{-1}_{t,t'}$
is proportional to $\d(t - t')$.  Since the bias voltage enters through the
respective lead equilibrium density matrix, see Eq.\
\eqref{eqn:StipulatedInitialState}, electronic energies are the same in both
leads. We have used the effective dot energy in Eq.\
\eqref{eqn:ContinuousAction:dot} defined as 
\begin{equation}\label{eqn:EffDotFrequency}
E_{\s} (t)=E_{\s}^+ (t)\equiv\e_{\s} + \frac{U}{2} + J \s \tau(t) + i \sigma
\zeta(t)
\lambda_{\dt}\, ,
\end{equation}
on the forward branch, while $E_{\s}^- (t) = E_{\s}^+ (t)^*$ on the backward branch. 

The polynomial $P[\{\tau\}]$ in Eq.~\eqref{eqn:KeldyshPathIntZ0} depends on the
impurity path $\{\tau\}=\{\tau^+\} (\{\tau^-\})$ for the forward (backward) branch of the contour.
Then, we collect all indices of the flips into the
tuple  $T_{\text{flip}}^{+} = (k^{+}_{m - \ell},\ldots,k^{+}_{1})$ (sorted in
ascending order) along
the forward path $\{\tau^+\} := (\tau_N,\ldots,\tau_1)$ with $\tau_{k^+} \ne
\tau_{k^+ - 1}$ for all $k^+ \in T_{\text{flip}}^+$. Accordingly, 
$T_{\text{flip}}^- = (k^-_{\ell},\ldots,k^-_{1})$ is the tuple of ascending flip
indices along the backward path $\{\tau^-\} := (\tau_{2N},\ldots,\tau_{N+1})$
with $\tau_{k^-} \ne \tau_{k^- + 1}$ for all $k^- \in T_{\text{flip}}^-$. Note
that a flip index on the backward path is labelled according to the
\emph{smaller} step index of the flipping spins corresponding to the
\emph{later} time. The impurity polynomial can be expressed in terms of 
the Grassmann fields as 
\begin{equation}\label{eqn:KeldyshFlipFlopPolynomial}
P [\{ \tau \}]:=\prod_{j \in T_{\text{flip}}^-}  \hrass{d}_{\tau_j}^{j +
1} \grass{d}_{-\tau_j}^{j} \prod_{k \in T_{\text{flip}}^+} 
\hrass{d}_{-\tau_k}^{k} \grass{d}_{\tau_k}^{k - 1}\, .
\end{equation}
Figure \ref{fig:ConstructP} illustrates an example of an impurity path. 
The fields of the forward and backward parts of the {\em continuous\/} inverse
Green's 
function $(G^{\text{el}})^{-1}$ are not coupled since the corresponding
Hamiltonian is diagonal. However, the associated {\em discrete\/} version
connects
fields from both branches via the upper right element $(1, 2 N)$, which is due
to the system's initial state
$\rho (t_i)$, see Ref.\ \cite{kamenev-2009-58} for details. Throughout this
work, we assume 
 factorising initial conditions 
\begin{equation}\label{eqn:StipulatedInitialState}
\rho (t_i) = \ket{0, \tau_i}\bra{0, \tau_i}\;\rho_{\text{L}}\rho_{\text{R}}\, ,
\end{equation}
where $\rho_{p} \propto \exp \{ -\b \sum_{\vec{k} \s} (\e_{\vec{k}} - \mu_{p})
c^\dagger_{\vec{k} \s p} c_{\vec{k} \s p} \}$ is 
the equilibrium density matrix of lead $p = \text{L}/\text{R}$ at temperature $T$ with $\b = (k_{\text{B}} T)^{-1}$ and $\ket{0,
\tau_i}$ denotes the
empty dot with the impurity in the initially prepared orientation
$\tau_i=|\uparrow\rangle$.  

When constructing the action $S_O [\eta]$ for an observable $O(t_m)$,
evaluated at the measurement time $t_m$, we replace every instance of an
impurity- or electron
operator in the observable by the corresponding spin- and Grassmann fields from 
the {\em forward\/} branch at the time step closest to $t_m$. 
This choice is arbitrary and  a replacement on the backward branch would not
change physical results.
Let us assume that $t_m - t_i = (k_m - 1) \dt$.  
The electron operators are replaced by their Grassmann field with time
index $k_m$.
Correspondingly, we substitute the Pauli matrices according to 
$\tau_\nu (t_{m}) \mapsto \tau_\nu^{(k_m)}$ with $\tau_x^{(k_m)} := (1 -
\tau_{k_m}
\tau_{{k_m}-1})/2, \tau_y^{(k_m)} := -i (\tau_{k_m} - \tau_{k_m-1})/2$ and
$\tau_z^{(k_m)} :=
\tau_{k_m}$. Since the matrix elements $\bra{\tau'} \tau_{x,y} \ket{\tau}$ are
non-zero only 
for $\tau \ne \tau'$, the Pauli matrices should be replaced by field
expressions that include neighbouring spins. In other words, only if a flip-flop
occurs at
time $t_m$, the fields $\tau_{x, y}^{(k_m)}$ are non-zero. On the forward
branch, a
flip-flop with $\tau_{k} = -\tau_{k-1}$ is associated with time step $k$.
Then,  $S_{O} [\eta]:= \eta O$ and 
\begin{equation}\label{eqn:PathIntObservable}
\expect{O} (t_{m})=-i \partial_{\eta} \ln \mathcal{Z} [\eta] \vert_{\eta=0}.
\end{equation}
A generalisation to observables with two and more time parameters, e.g.,
correlation functions, is possible via higher-order derivatives of the
generating function.
We calculate the charge current $I(t_m)$ via the source term
\begin{equation}\label{eqn:CurrentSourceTerm}
S_I=-\frac{i e \eta}{2} \sum_{\vec{k} \s p}
p \left( \g \hrass{d}_{\s}^{(k_m)} \grass{c}_{\vec{k}  \s
p}^{(k_m)} - \g^* \hrass{c}_{\vec{k} \s p}^{(k_m)} \grass{d}_{\s}^{(k_m)}
\right) \, ,
\end{equation}
and the expectation value of the impurity $\langle \tau_z(t_m)\rangle$ from
$S_{\tau_z}= \eta \tau_{k_m}$.

\subsection{Tracing Out Electron Degrees of Freedom}
\label{trace_out}
Next, we perform traces over the lead degrees of freedom and perform the path
integral over all Grassmann
fields $\hrass{c}_{\vec{k}\s p}, \grass{c}_{\vec{k}\s p}$. Contour time
integrations are transferred to their respective real time counterparts. This
results in  
two integrals over real time for the ($+$) and ($-$) branch and generates the
$2\times2$-matrix structure for the Keldysh Green's function.
The resulting generating function is written as a path integral with the
effective electronic dot action
$S_{\text{el}} = S_{\text{el}}^{\text{dot}} + S_{\text{env}}$ with the
(effective) environmental action  
\begin{equation}
\label{eqn:SEnv:S}
S_{\text{env}}=\sum_{\s p}
\int_{-\infty}^{\infty}\hspace*{-4mm} dt
\int_{-\infty}^{\infty}\hspace*{-4mm} dt'\,
\hrass{d}_{\s}(t)\boldsymbol{\g} (p, t - t')
\Bigl\{1+\frac{i e \eta p}{2}[\d_{m} (t') - \d_{m} (t)]
\Bigr\}\grass{d}_{\s} (t') \, .
\end{equation}
Here, we have introduced the time non-local Keldysh matrix
\begin{equation}\label{eqn:gamma:time}
\boldsymbol{\g} (p, t - t')
=\frac{\G}{2 \b}
\frac{e^{-i \mu_p (t - t')}}{ \sinh [ \pi (t - t')/\b ] }
\begin{pmatrix}
-1&1\\
1&-1\end{pmatrix}\, .
\end{equation}
This equation only holds for $t-t'\ne 0$. The singularity 
for $t=t'$ will be addressed below.
The exponential decay of $\boldsymbol{\g}$ for $|t-t'| \to \infty$ is the
cornerstone of the ISPI method, which allows us to truncate certain long-time correlations (see below).
Equation (\ref{eqn:SEnv:S}) can also be given in terms of an environmental
Green's function
\begin{equation}
\label{eqn:SEnv:2}
S_{\text{env}}=
\sum_{\s}
\int_{-\infty}^{\infty}\hspace*{-4mm} dt
\int_{-\infty}^{\infty}\hspace*{-4mm} dt'\,
\hrass{d}_{\s} (t)
\bigl( G_{\text{env}} \bigr)^{-1}_{t,t'} \grass{d}_{\s} (t')
\text{~~with~~}
\bigl( G_{\text{env}} \bigr)^{-1}_{t,t'}
=\bigl( G_{\text{env}}^0 \bigr)^{-1}_{t,t'}
+\eta \bigl( G_{\text{env}}^I \bigr)^{-1}_{t,t'}
\end{equation}
and
\begin{equation}\label{eqn:GEnv:I}
\bigl( G_{\text{env}}^I \bigr)^{-1}_{t,t'}
= \frac{e \G}{2 \b} \frac{\sin [ e V (t - t') / 2 ] }{ \sinh  [\pi (t - t')/\b ] }
\begin{pmatrix}
\d_{m} (t) - \d_{m} (t')
&-\d_{m} (t)
\\ \d_{m} (t')
&0\end{pmatrix} \, .
\end{equation}
This expression is well defined for all values of
$t$ and $t'$. An explicit expression for $(G^0_{\text{env}})_{t,t'}$ is given
below in Eq.~\eqref{G0envttp}. 

\begin{figure}[t!]
\begin{center}
\includegraphics{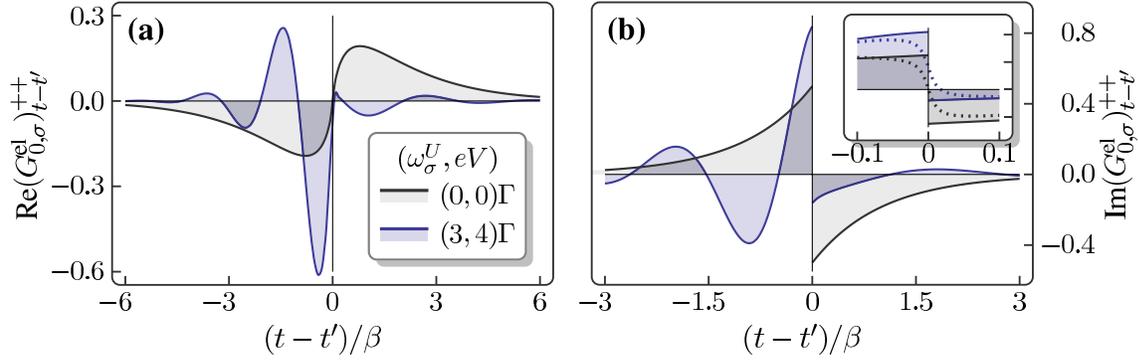}
\end{center}
\caption{
\label{fig:FreeGF}
$(G_{0,\s}^{\text{el}})^{++}$ vs.\ time difference $t - t'$ for two
combinations of $\w_{\s}^{U}$ and $e V$ for $\beta\G =
1/5$.  Except for $t=t'$, the function is smooth for all real times. 
 The imaginary part is discontinuous at $ t = t'$, which turns out
to be harmless for our numerical scheme. A possible way to cure it,  is to
introduce a high frequency cutoff $\exp\{- \abs{\w}/\w_{c} \}$ in Eq.\ 
\eqref{eqn:FourierSpaceFreeGF}, see dotted lines in the inset
for the same vertical scale for $\w_{c} = 100 \G$.}
\end{figure}

Before we proceed, we address the $(t - t')^{-1}$ singularity of 
$(G_{\text{env}}^0)^{-1}$.  In contrast to 
its inverse, the Green's function $G_{\text{env}}^0$ itself is finite at $t =
t'$. Still, matrix elements decay exponentially
with growing time differences.  The calculation of observables does not suffer
from this singularity either, since it cancels in the fraction 
\begin{equation}\label{eqn:CancelDivergence}
\frac{\mathcal{Z}[\eta] - \mathcal{Z}[0]}{\eta \mathcal{Z}[0]}  \approx
i \expect{O} (t_{m}),
\end{equation}
which is used to numerically evaluate $\expect{O}$. Since the divergence
does neither depend on the paths of the impurity- and the HS-spins nor on
$\eta$,  we may rescale $\mathcal{Z}[\eta]$ by the singular factor without
affecting $\expect{O} (t_{m})$. For clarity, we keep the same notation for the
rescaled generating function. The singularity originates from 
$(G_{\text{env}}^0)^{-1}$. We collect all non-interacting contributions, 
which do not depend on $\eta$ into the 
sum $\sum_{\s} (G_{0,\s}^{\text{el}})^{-1}_{t, t'}
=\sum_{\s}(G_{\text{dot},\s}^{\text{el},0})^{-1}_{t, t'} +
(G_{\text{env},\s}^{0})^{-1}_{t, t'}$ with 
\begin{eqnarray}
(G_{\text{env},\s}^{0})^{-1}_{t, t'}
=\sum_{p} \boldsymbol{\g} (p, t - t'), \nonumber \\
(G_{\text{dot},\s}^{\text{el},0})^{-1}_{t, t'}
=\d (t - t')\begin{pmatrix}
i \partial_t - \w_{\s}^U&0\\
0&-i \partial_t + \w_{\s}^U
\end{pmatrix}\, , 
\label{G0envttp}
\end{eqnarray}
with $\w_{\s}^{U} := (\e_{\s} + U /2 )$.  The 
Fourier transform of $(G_{0,\s}^{\text{el}})^{-1}_{t,
t'}$  is obtained as  
\begin{equation}\label{eqn:FourierSpaceFreeGF}
(G_{0,\s}^{\text{el}})^{-1}_{\w, \w'} = 2 \pi  \d (\w - \w')
\begin{pmatrix}
\w - \w_{\s}^U + i \G [F(\w) - 1]&-i \G F(\w)\\
i \G [2 - F(\w)]&-\w + \w_{\s}^U  + i \G [F(\w) - 1]
\end{pmatrix}\,.
\end{equation}
where we have introduced $F(\w)= f_{\text{L}} (\w) +
f_{\text{R}} (\w)$ as the sum of the two lead Fermi distributions. This 
 $2\times 2$-matrix is inverted algebraically and transformed back into
time space by complex contour integration \cite{Thesis_Becker11} or numerical
integration. The function $(G_{0,\s}^{\text{el}})^{++}(t-t')$ is shown in Fig.\
\ref{fig:FreeGF}. 
 Whenever this divergence arises, we will remove it, see for instance in Eq.\ 
\eqref{eqn:FinalGenFunc},
by multiplication of $-i \det
G_{0,\s}^{\text{el}}$, or equivalently, we replace $i(G^{\text{eff}}_{\s})^{-1}$
with  
\begin{equation}\label{eqn:DMatrix1}
D_{\s} [\eta]
:=G_{0,\s}^{\text{el}} (G^{\text{eff}}_{\s})^{-1}
=\mathbbm{1} + G_{0,\s}^{\text{el}} (\s \Sigma_{\s}^0 +  \eta
\Sigma_{\s}^{\eta})\, ,
\end{equation}
with the self-energy
\begin{equation}\label{eqn:DMatrix2}
(\Sigma_{\s}^0)_{k l}=\d_{k l}\begin{pmatrix}
-J \tau_k - i \zeta_k \lambda (\dt)&0\\
0& J \tau_k - i \zeta_k \lambda^* (\dt)
\end{pmatrix}\dt \, .
\end{equation}
The particular form of $\Sigma_{\s}^{\eta}$ depends on the observable $O$. When
$O = I$, we identify it with $(G_{\text{env}}^I)^{-1}$ 
of Eq.\ \eqref{eqn:GEnv:I} and otherwise with $(G^O)^{-1}$. 

The discrete form of the matrix $D_{\s}$ follows via the relation  
\begin{equation}\label{eqn:FreeGF:Discrete}
		(G_{0, \s}^{\text{el}})_{k l}
	=	\frac{1}{\dt^{2}}
		\int_{t_k -\dt/2}^{t_k + \dt/2}
\hspace*{-10mm}dt\hspace*{5mm}\int_{t_l-\dt/2}^{t_l + \dt/2}
\hspace*{-10mm}dt'\hspace*{2mm}
			(G_{0,\s}^{\text{el}})_{t, t'}
	\approx
		(G_{0,\s}^{\text{el}})_{t_k, t_l}
	\quad\text{with}\quad
		t_k = t_1 + (k - 1) \dt \, .
\end{equation}
To remove the singularity at $k=l$, we choose the regularisation 
\begin{equation}\label{eqn:RepairDiscontinuity}
		(G_{0, \s}^{\text{el}})_{l l}
	=	[(G_{0,\s}^{\text{el}})_{(t - t') \to0^-}+(G_{0,
\s}^{\text{el}})_{(t - t') \to0^+}]/2\, .
\end{equation}
 Alternatively, introducing a high frequency cut off $\exp\{-\abs{\w}/\w_{c} \}$
in the Green's function
 \cite{weiss:195316} yields a consistent result, see the inset of
Fig.~\ref{fig:FreeGF}(b). 
For $\w_{\s}^{U} = 3 \G$ and $e V = 4 \G$, Eq.\  
\eqref{eqn:RepairDiscontinuity} yields $(G_{0,\s}^{\text{el}})_{l, l} \approx
0.3384 i$ and with the cutoff method with $\w_{c}=100 \G$, we obtain $0.3245
i$ (the difference of $\sim 4\%$ decreases for larger $\w_{c}$). However, using
Eq.\ \eqref{eqn:RepairDiscontinuity} has two advantages: first, it
does not modify off-diagonal Green's matrix elements, and, second, we do not
need 
an additional parameter $\w_{c}$. 

We emphasize that both methods of regularisation obey the necessary causality
relation, $(G_{0, \s}^{\text{el}})_{l l'}^{++} + (G_{0, \s}^{\text{el}})_{l
l'}^{--} - (G_{0, \s}^{\text{el}})_{l l'}^{+-} - (G_{0, \s}^{\text{el}})_{l
l'}^{-+} = 0$. This follows from the causality structure of the Green's matrix
and the self-energies $\Sigma$ \cite{kamenev-2009-58}. Here, the diagonal
elements of both $G_{0, \s}^{\text{el}}$ and the $\Sigma$ matrices have to be
understood as the average, i.e., the \emph{integral} of the time non-local
matrix elements in an interval $\dt$ \emph{around the point $t-t'=0$}. The
discrete version of  $\Sigma_{\s}^{\eta}$
follows accordingly. For the current, e.g., using Eq.\
\eqref{eqn:GEnv:I} yields 
\begin{equation}\label{eqn:DiscreteSigmaI}
\bigl( \Sigma_{\s}^{\eta} \bigr)_{k l}=\frac{e \G \dt}{2 \b}
\frac{\sin [ e V (k - l) \dt / 2 ] }{ \sinh [  \pi (k- l) \dt/\b ] }
\begin{pmatrix}\d_{k,m} - \d_{l,m}&-\d_{k,m}\\
\d_{l,m}&0\end{pmatrix} \, .
\end{equation}

We note that also source terms for observables $O$ containing dot fields may be 
added to the action. 
The effective full inverse Green's function
$(G^{\text{eff}})^{-1}$ is given either by 
$(G_{\text{dot}}^{\text{el}})^{-1} +
(G_{\text{env}})^{-1}$ for the current or $(G_{\text{dot}}^{\text{el}})^{-1} +
\eta (G^O)^{-1} + (G_{\text{env}}^0)^{-1}$ for other observables. 
 Plugging in those pieces, the remaining path integral is recast as a discrete
sum, i.e., 
\begin{eqnarray}\label{eqn:KeldyshPathIntZ0poly}
\mathcal{Z}[\eta]&=&\sum_{\{\tau, \zeta\}}
\int\!\! \mathcal{D} [ \hrass{d}_{\s} \grass{d}_{\s} ] P [\{\tau\}] \exp
\Bigl\{
i
\overbrace
{
	\sum_{\s}
	\int_{-\infty}^{\infty}\hspace*{-4.5mm} \dd t
	\int_{-\infty}^{\infty}\hspace*{-4.5mm} \dd t'\;
	\hrass{d}_{\s} (t)
	\bigl( G_{\s}^{\text{eff}} \bigr)^{-1}_{t,t}
	\grass{d}_{\s} (t')
}^{\displaystyle S_{\text{eff}}}
\Bigr\} \nonumber\\
&=&\sum_{\{\tau, \zeta\}} \langle P[\{\tau\}]\rangle \prod_{\s} \det \{ ( i
G_{\s}^{\text{eff}} [\{\tau,\zeta\},\eta])^{-1}\}.
\end{eqnarray}
Hence, we have obtained the Keldysh partition function as a
sum over expectation values of the polynomial $P$ of Grassmann numbers in a
system with Green's function
$G_{\s}^{\text{eff}}$. The next step is to derive an expression for $\expect{P
[\{\tau\}] }$ that refers 
to $G_{\s}^{\text{eff}}$ only by applying  Wick's theorem
\cite{kamenev-2009-58}. 
  With Eq.\ \eqref{eqn:KeldyshFlipFlopPolynomial}, we have
\begin{equation}\label{eqn:ExpVal:PK}
\expect{P [\{\tau\}] } =
\expect
{
	\prod_{j \in T_{\text{flip}}^-} \hrass{d}_{\tau_j}^{j + 1} \grass{d}_{-\tau_j}^{j}
	\prod_{k \in T_{\text{flip}}^+} \hrass{d}_{-\tau_k}^{k} \grass{d}_{\tau_k}^{k - 1}
}.
\end{equation}
For an odd number $m$ of flip-flops the expectation value vanishes, since each
process contributes a
creator and an annihilator for electrons with opposite spins. 
Odd $m$ implies an odd number of alternating products of creators and
annihilators, 
the number of $\hrass{d}$ is different from the number of $\grass{d}$. When
applied to any state 
$\ket{\psi}$ in the trace, this 
 changes the particle count by $1$ and the projection with $\bra{\psi}$
vanishes. Therefore we
have to consider paths with even $m$ only.
\begin{figure}
\begin{center}
\includegraphics{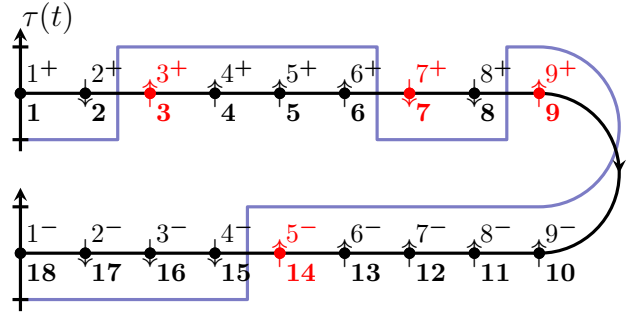}
\end{center}
\caption{
\label{fig:ExampleTau}
Exemplary impurity path (blue) with $N = 9$ and $m=4$ flip-flops along the
Keldysh contour.}
\end{figure}
As an example, we evaluate Eq.\ \eqref{eqn:ExpVal:PK} for the
impurity path shown in Fig.\ \ref{fig:ExampleTau} before we turn to the general
formalism and arbitrary paths $\{ \tau \}$.
The exemplary path features $m=4$ flip-flops and we find
\begin{equation}\label{eqn:ExpValPK:Example}
\begin{split}
\expect{P}
= \expect
{
	\hrass{d}_{\uparrow}^{\,15}
	\grass{d}_{\downarrow}^{14}
	\hrass{d}_{\downarrow}^{\,9}
	\grass{d}_{\uparrow}^{8}
	\hrass{d}_{\uparrow}^{\,7}
	\grass{d}_{\downarrow}^{6}
	\hrass{d}_{\downarrow}^{\,3}
	\grass{d}_{\uparrow}^{2}
} 
=\underbrace
{
\displaystyle
\expect
{
	\hrass{d}_{\uparrow}^{\,15}
	\grass{d}_{\uparrow}^{8}
	\hrass{d}_{\uparrow}^{\,7}
	\grass{d}_{\uparrow}^{2}
}
}_{\displaystyle P^{\uparrow}}\;
\underbrace
{
\displaystyle
\expect{
	\grass{d}_{\downarrow}^{14}
	\hrass{d}_{\downarrow}^{\,9}
	\grass{d}_{\downarrow}^{6}
	\hrass{d}_{\downarrow}^{\,3}
}
}_{\displaystyle P^{\downarrow}}.
\end{split}
\end{equation}
$S_{\text{eff}}$ and the expectation value of the mixed operator product
factorise with respect to the spin degree of
freedom. Even in the presence of spin-mixing terms this factorisation remains
valid.
Applying Wick's theorem to Eq.\ \eqref{eqn:ExpValPK:Example} yields 
$\expect{P^{\sigma}}\equiv -\det\,\Xi_{\sigma}$ with 

\begin{equation}\label{eqn:ExpXi}
\expect{P^{\uparrow}}=-\det\!\!
\underbrace
{
\begin{pmatrix}
(G_{\uparrow}^{\text{eff}})_{2, 7}&
(G_{\uparrow}^{\text{eff}})_{2, 15}\\
(G_{\uparrow}^{\text{eff}})_{8,7}&
(G_{\uparrow}^{\text{eff}})_{8, 15}
\end{pmatrix}
}_{\displaystyle \phantom{_{\uparrow}}\Xi_{\uparrow}}
\text{ and }
\expect{P^{\downarrow}}
=-\det\!\!
\underbrace
{
\begin{pmatrix}
(G_{\downarrow}^{\text{eff}})_{6, 3} &
(G_{\downarrow}^{\text{eff}})_{6, 9}\\
(G_{\downarrow}^{\text{eff}})_{14, 3}&
(G_{\downarrow}^{\text{eff}})_{14,9}
\end{pmatrix}
}_{\displaystyle \phantom{_{\downarrow}}\Xi_{\downarrow}}.
\end{equation}
The general procedure to construct $\Xi_{\s}$ for an arbitrary path $\{\tau \}$
having an even number $m$ of 
flip-flops is as follows,
\begin{enumerate}
\item	\label{enum:ContructXi:First}
Construct the flip-flop polynomial $P^{\s} [\{ \tau\}]$.
\item	\label{enum:ContructXi:Rows}
Assign indices $q_1 < \ldots < q_{m/2}$ of 
\emph{annihilator} fields $\grass{d}_{\s}^{q_1}, \ldots,
\grass{d}_{\s}^{q_{m/2}}$ that appear in $P^{\s} [\{ \tau \}]$ to the
\emph{rows} of $\Xi_{\s}$.
\item	\label{enum:ContructXi:Cols}
Assign indices $r_1 < \ldots < r_{m/2}$ of 
\emph{creator} fields $\hrass{d}_{\s}^{r_1}, \ldots, \hrass{d}_{\s}^{r_{m/2}}$
that appear in  $P^{\s} [\{ \tau \}]$ to the \emph{columns} of $\Xi_{\s}$.
\end{enumerate}
Using matrix element $(\Xi_{\s})_{k, l} =  -i
\expect{\grass{d}_{\s}^{q_k} \hrass{d}_{\s}^{r_l}} = (G_{ \s}^{\text{eff}})_{q_k
r_l},$ the final expression for the generating function follows as 
\begin{equation}\label{eqn:FinalGenFunc}
		\mathcal{Z}[\eta]
	=	\lim_{\dt \to 0} \sum_{\{\tau, \zeta\}}
		(-1)^{\ell} \Bigl(\frac{J \dt}{2 }\Bigr)^{m}
		\!\!\!\exp \{ i S_{\text{imp}} \}
		\prod_{\s}\det i ( G_{\s}^{\text{eff}} )^{-1}\det \Xi_{\s},
\end{equation}
where the summation over impurity paths is restricted to tuples $\{\tau\}$  with
$\tau_1 = \tau_{2 N} = \tau_i$, i.e., correct boundary conditions along the
Keldysh contour are imposed. The limit $\delta_t\to 0$ appears explicitly here,
since there is no continuous measure used for the discrete spin paths, neither
for the HS- nor for the impurity spins.
\begin{figure}
\begin{center}
\includegraphics{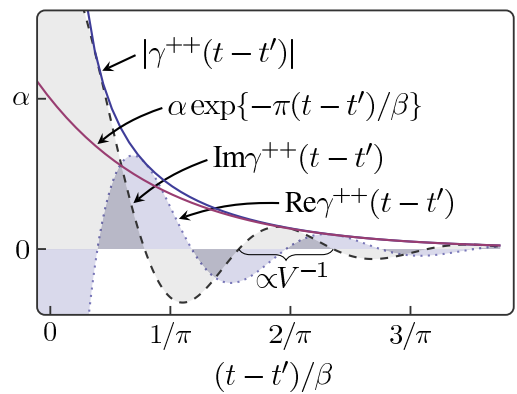}
\caption
{
\label{fig:GammaDecay}
Dependence of $\g^{++}$ on $t-t'$ for  $\alpha = 
\G/(2\b)$ and $e V \b = 4\pi$.  We depict the absolute value and the real
as well as the imaginary part.
For $\pi |\Delta t|/\b \gtrapprox 2$, the absolute value decays exponentially 
with a decay time proportional to $T^{-1}$. The period of the oscillations in
the real and imaginary parts is determined by $V$.}
\end{center}
\end{figure}

\subsection{Iterative summation of the path integral}
 \label{subsec:CoherencApprox}

The exact generating function in Eq.~\eqref{eqn:FinalGenFunc} is intractable due
to the exponentially growing size of the matrices for long propagation times and an
adept numerical treatment is necessary to proceed. 
The off-diagonal elements of $(G^{\text{eff}})^{-1}$, given in terms of 
$\boldsymbol{\g} (p, t - t')$ in Eq.\ \eqref{eqn:gamma:time} decay exponentially
 with increasing distance $t - t'$ from the diagonal,
at finite $T$ and/or $V$ \cite{weiss:195316}. This fact is illustrated in Fig.\
\ref{fig:GammaDecay}. In
addition, the bias voltage induces oscillations.  The correlation or  memory
time $\TauC=: (K - 1) \dt$ determines the range of the lead-induced
correlations, they are exponentially suppressed for time differences larger
than 
$\TauC$. This allows us to restrict the path summation to a finite memory time
window and thus, 
the number of paths that contribute to $\mathcal{Z}[\eta]$, originating from the
magnetic impurity as well as from the Coulomb interaction, is drastically
reduced. We use $K$ as a memory time 
parameter in the ISPI
scheme in an extrapolation procedure for 
$\TauC \to \infty$. The latter eventually gives converged results independent of
$\TauC$.

To proceed, we rotate the basis unitarily, thereby rearranging the matrix
elements of $(G^{\text{eff}}_{\s})^{-1}$
such that increasing distances with respect to the diagonal correspond to
increasing time differences.
For a given $K$, we obtain the block band-matrix
\be\label{eqn:GFinKBlocks}
 (G^{\text{eff}}_{\s})^{-1}\approx \left(
    \begin{array}{cccc}
      \boxx{A}{1,1}{\sigma}&\boxx{A}{1,2}{\sigma}&  & \\[1mm]
      \boxx{A}{2,1}{\sigma}&\boxx{A}{2,2}{\sigma}& \ddots  & \\[1mm]
      & \ddots & \ddots & \hspace{2ex}\boxx{A}{N_c-1, N_c}{\sigma} \\[1mm]
      &  & \hspace{2ex}\boxx{A}{N_c, N_c-1}{\sigma} & \boxx{A}{N_c, N_c}{\sigma} \\[1mm]
   \end{array}
  \right)
\ee
where $N_C = N / K$ and the blocks $\boxx{A}{k,l}{\s}$ are $K-$dimensional
matrices, whose entries are given by those of $(G^{\text{eff}}_{\s})^{-1}$
taken from the rows and columns in the range of $\{ (j- 1) K + 1, \ldots, j K\}$
with $j = k, l$.  Since we neglect exponentially small components, the
$\boxx{A}{}{}$-blocks in 
the upper (lower) secondary diagonal have an upper (lower) triangular
structure. 
The $\Xi_{\s}$ matrices in Eq.\ \eqref{eqn:FinalGenFunc} naturally inherit 
the same block structure with the
modification that the corresponding blocks $\boxx{B}{k,l}{\s}$ are in general 
not quadratic. Their dimensions are determined by the number of flip-flops
within
the respective time interval.

\begin{figure}
\begin{center}
\includegraphics{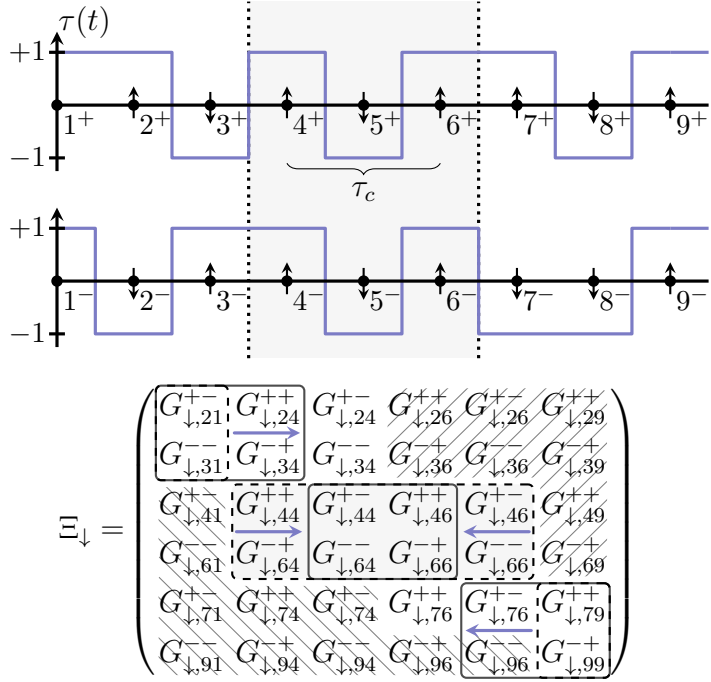}
\caption
{
\label{fig:ApproxXiMatrix}
 An example of an impurity path with 12 flip-flops (top) and the
associated $\Xi_{\downarrow}$-matrix (bottom) for $\TauC = 2
\dt$ ($K = 3$). The discrete path (arrows on the vertices)  
has a length of $8\dt$ ($N = 9$) and is divided into $N_C = 3$ segments of
length $K$ (separated by dotted lines). Depending on the flip distribution, 
the diagonal blocks $\boxx{B}{i,i}{\downarrow}$ are in general not
quadratic (boxes with dashed frames) and their determinants do not exist.  To
render them quadratic, we reassign the fields of the flips closest to the
segment borders (blue arrows). The hatched elements are exponentially small and
neglected.}
\end{center}
\end{figure}

To illustrate the scheme, a particular impurity path is shown in Fig.\ 
\ref{fig:ApproxXiMatrix}, which is of length $8 \dt$, consisting of 
$2 N = 18$ vertices having $12$ flip-flops. The calculation of the determinant
in our scheme needs 
quadratic block matrices on the diagonal, which we obtain
 again by a unitary transformation. The off-diagonal blocks are reshaped
accordingly. In Fig.~\ref{fig:ApproxXiMatrix} this procedure 
is illustrated for 
$\Xi_{\downarrow}$ with $\TauC = 2 \dt (K = 3)$, obtained after the
rearrangement. The hatched matrix elements are disregarded. The path with $N =
9$ is divided into $N_C = 3$ 
 segments with $K$ vertices on each branch. In analogy to the blocks
$\boxx{A}{i,i}{\downarrow}$, the diagonal blocks
$\boxx{B}{i,i}{\downarrow}$ contain all matrix elements
$G^{\alpha}_{\downarrow, q r}$ with $(i - 1) K < q, r \le i K $ (dashed boxes).
Since the number of flip-flops is odd,  all the matrices
$\boxx{B}{i,i}{\downarrow}$ are not quadratic. Instead, this particular case
yields 
 $2\times1$-matrices $\boxx{B}{1,1}{\downarrow}$ and
$\boxx{B}{3,3}{\downarrow}$, and a 
$2\times4$-matrix $\boxx{B}{2,2}{\downarrow}$. The quadratic blocks
required for the iteration scheme are obtained by reassigning the earliest and
latest creator fields $\hrass{d}_{\downarrow, 4}^+$ and $\hrass{d}_{\downarrow,
6}^-$ from the second segment to the first and third, respectively (blue
arrows). Such a reordering is always possible and renders all diagonal
blocks of $\Xi_{\s}$ quadratic. 

We define the recursive notation $\mathbf{X}=A, B$ to compactify the computation
of the determinant  for the blocked Keldysh partition function as 
\begin{equation}\label{eqn:BlockMatrixX}
\mathbf{X} = 
\dboxx{\mathbf{X}}{D} \text{\hspace{1ex} with \hspace{1ex}} 
\dboxx{\mathbf{X}}{D} =  \left(
    \begin{array}{cc}
      {\boxx{X}{i,i}{}\hspace{2ex}}&\boxx{X}{i+1,i}{} \\[1mm]
      \boxx{X}{i,i+1}{}&\dboxx{X}{D-1}  \\[1mm]
    \end{array}
  \right)
\text{\hspace{1ex} and \hspace{1ex}} 
\dboxx{\mathbf{X}}{1} =  \boxx{X}{D,D}{} \, .
\end{equation}
The double line denotes matrices which themselves consists of blocks with 
the subscript $D$ giving their dimension in blocks. The determinant of
$\mathbf{X}$
is calculated iteratively \cite{weiss:195316} in $D-1$ steps of which each
performs the following manipulations: 
\begin{enumerate}
\item Perform a Gaussian elimination of 
the block in the second row, first column.
\item Neglect in the resulting element of second row, second  column
 products like $\boxx{X}{k-1,k}{}\, \boxx{X}{k,k + 1}{}$, which connect 
segments beyond the nearest neighbour.
\item Expand the determinant after the first column,  thus
reducing the problem by one in block dimension.
\end{enumerate}
While step (i) and (iii) are exact algebraic operations, 
step (ii) is the second building block of the ISPI method, necessary for the
scheme to remain consistent with neglecting
correlations beyond $\TauC$. A step $k \to k + 1$ is performed solely
based on the determinant after step $k$ and the spin orientations 
in segments $k$ and $k + 1$. Here we stress that within the time window $\TauC$,
the ISPI scheme takes into account important non-Markovian effects in a natural
way. 
In the present work, these correlations are lead-induced. Within a typical
Markovian approximation,  the real-time dependence of the GF in
Fig.~\ref{fig:ApproxXiMatrix} is replaced by $\delta(t-t')$. Including terms in
the iteration that connect
segments beyond nearest neighbouring $K$-blocks would require information about
impurity spins in ``earlier'' segments $< k$, which are beyond $\TauC$. 
After step (iii), we arrive at
\begin{equation}\label{eqn:FirstIteration}
\det \mathbf{X}= \det \boxx{X}{1,1}{} \left(
    \begin{array}{cc}
      \boxx{X}{2,2}{\prime}&\boxx{X}{2,3}{} \\[1mm]
      \boxx{X}{3,2}{}&\dboxx{\mathbf{X}}{D-2}  \\[1mm]
    \end{array}
  \right) \, ,
\end{equation}
where $\boxx{X}{2,2}{\prime} = \boxx{X}{2,2}{} - \boxx{X}{2,1}{}
\boxx{X}{1,1}{-1}
\boxx{X}{1,2}{}$. Subsequent iteration gives the final relation 
\begin{equation}\label{eqn:FinalIterativeMatrix}
\det \mathbf{X}=\det \boxx{X}{1,1}{}\prod_{i = 2}^D
\det\{
{\boxx{X}{i, i}{}-\boxx{X}{i,i - 1}{} \boxx{X}{i- 1,i - 1}{-1} \boxx{X}{i -
1,i}{}
}\}
\end{equation}

Next, we construct the $\Xi_{\s}$-matrices for finite correlation
times starting from Eq.\ \eqref{eqn:FinalGenFunc} by filling the
entries with the elements of $G^{\text{eff}}_{\s} 
[\{\tau, \zeta\}]$. The latter depends on the entire spin path $\{\tau, \zeta\}$
between $t_i$ and $t_f$. In principle, inversion of the full inverse Green's 
function is possible but out of reach for practical applications since the 
numerical effort grows exponentially with propagation time. To
remain consistent with the finite correlation time approach, we have to find
approximations of $\boxx{B}{i,i (\pm 1)}{\s}$ that depend on 
spins of the nearest neighbour segments $i (\pm 1)$. We observe that
blocks on the secondary diagonals contribute much less than those on the main
diagonal. Hence, we expand $(G^{\text{eff}}_{\s})^{-1}$ in powers of the
off-diagonal blocks, yielding
\begin{equation}\label{eqn:ApproxGF}
	\boxx{G}{k,k}{} \approx \boxx{A}{k,k}{-1} 
	\quad\text{and}\quad
	\boxx{G}{k,l}{} \approx - \boxx{A}{k,k}{-1} \boxx{A}{k,l}{}
\boxx{A}{k,l}{-1}
\end{equation}
for all $1 \le k, l \le N_C$ and $\abs{k - l} = 1$ (the index $\sigma$ was
omitted). The blocks $\boxx{G}{k,l}{}$ are defined in analogy to the
$\boxx{A}{}{}$-blocks and the $\boxx{B}{}{}$-blocks are filled
as described above.

\begin{figure}
\begin{center}
\includegraphics{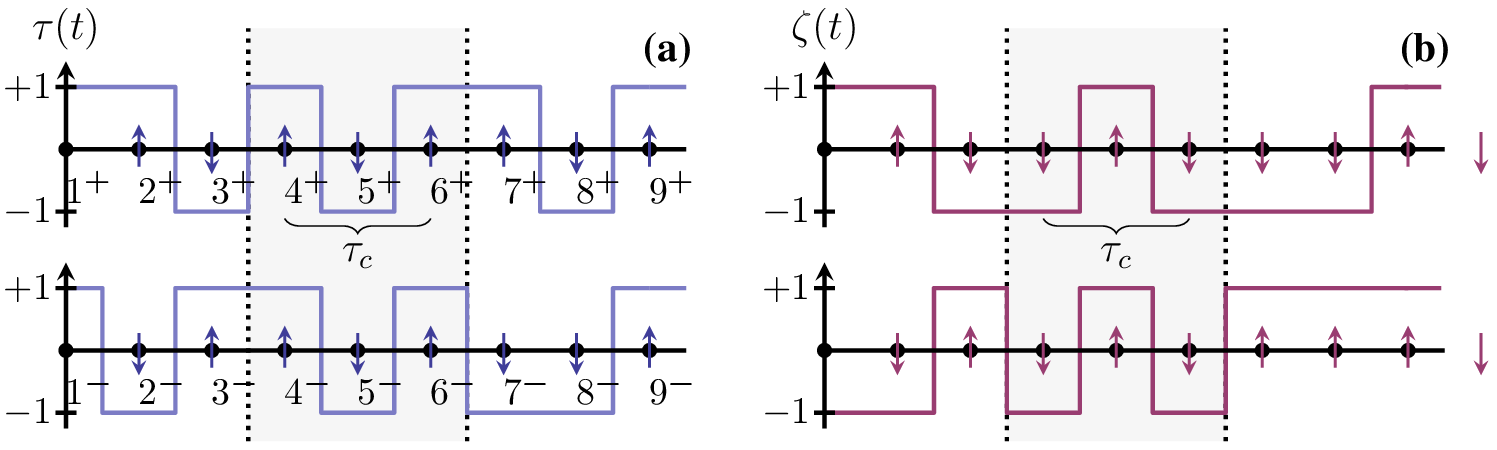}
\end{center}
\caption{\def\RedUp{{\color{plot_b} \boldsymbol{\uparrow}}}
\def\RedDown{{\color{plot_b} \boldsymbol{\downarrow}}}
\def\BlueUp{{\color{plot_a} \boldsymbol{\uparrow}}}
\def\BlueDown{{\color{plot_a} \boldsymbol{\downarrow}}}
\label{fig:SpinPathSegmentation}
Exemplary paths for the impurity (a) and the HS-field (b). 
For $\TauC = 2 \dt$ ($K = 3$) both are decomposed into segments of
length $K = 3$ in real time, which contain $2 K = 6$ spins. For example,
path segment $\{\tau \}_1 = (\tau_3^-, \tau_3^+, \tau_2^-,
\tau_2^+, \tau_1^-, \tau_1^+) = ({\color{plot_a} \boldsymbol{\uparrow}},
{\color{plot_a} \boldsymbol{\downarrow}}, {\color{plot_a}
\boldsymbol{\downarrow}}, {\color{plot_a} \boldsymbol{\uparrow}},
{\color{plot_a} \boldsymbol{\uparrow}}, {\color{plot_a}
\boldsymbol{\uparrow}})$.
Accordingly, we find that $\{\zeta \}_3 = (\zeta_9^-, \zeta_9^+, \zeta_8^-,
\zeta_8^+, \zeta_7^-, \zeta_7^+) = ({\color{plot_b} \boldsymbol{\uparrow}},
{\color{plot_b} \boldsymbol{\uparrow}}, {\color{plot_b} \boldsymbol{\uparrow}},
{\color{plot_b} \boldsymbol{\downarrow}}, {\color{plot_b}
\boldsymbol{\uparrow}}, {\color{plot_b} \boldsymbol{\downarrow}})$. The compact
notation 
combines the impurity- and HS spins as $\{\tau, \zeta \}_j$, e.g., $\{\tau,
\zeta\}_2 = (\BlueUp, \BlueUp, \BlueDown, \BlueDown, \BlueUp, \BlueUp, \RedDown,
\RedDown, \RedUp, \RedUp, \RedDown, \RedDown)$.}
\end{figure}

 Next, we use from Eq.~\eqref{eqn:DMatrix1} the relation $G^{\text{eff}}_{\s} =
D_{\s}^{-1} G_{0,\s}^{\text{el}}$, see Sec.~\ref{trace_out}.
The $\Xi$-matrices are free of any singularity as well and from
Eq.\ \eqref{eqn:ApproxGF}, we obtain the approximate inverse of $D_{\s}$. We
multiply the result with the free Green's matrix in block form. For step $k - 1
\to k$, we obtain 
\begin{eqnarray}\label{eqn:ApproxGF2}
		\boxx{G}{k - 1, k - 1}{}
	&=&\;
		\boxx{D}{k - 1, k - 1}{-1}
		\left\{
				\boxx{G_0}{k - 1, k - 1}{}
			-	\boxx{D}{k - 1, k}{}\boxx{D}{k, k}{-1}
\boxx{G_0}{k, k - 1}{}
		\right\}
\nonumber \\
		\boxx{G}{k -1, k}{}
	&=&\;
		\boxx{D}{k -1, k -1}{-1}
		\left\{
				\boxx{G_0}{k -1, k}{}
			-	\boxx{D}{k -1, k}{}\boxx{D}{k, k}{-1}
\boxx{G_0}{k, k}{}
		\right\}
\nonumber \\
		\boxx{G}{k, k - 1}{}
	&=&\;
		\boxx{D}{k, k}{-1}
		\left\{
				\boxx{G_0}{k, k -1}{}
			-	\boxx{D}{k, k - 1}{}\boxx{D}{k - 1, k - 1}{-1}
\boxx{G_0}{k - 1, k - 1}{}
		\right\}
\nonumber \\
		\boxx{G}{k, k}{}
	&=&\;
		\boxx{D}{k, k}{-1}
		\left\{
			-	\boxx{D}{k, k - 1}{}\boxx{D}{k - 1, k - 1}{-1}
\boxx{G_0}{k - 1, k}{}
		\right\}\, .
\end{eqnarray}
This allows us to construct the $\boxx{B}{}{}$-blocks without calculating the
inverse Green's function explicitly. 

Collecting all parts, we can finally express $\mathcal{Z} [\eta]$ iteratively as
\begin{equation}\label{eqn:FinalIterativeGenF:main}
\mathcal{Z}[\eta] =\sum_{
\{  \tau, \zeta \}_{N_C}} \mathcal{Z}_{N_C}
\text{~~~where~~~}
\mathcal{Z}_{j}=\sum_{
\{ \tau, \zeta \}_{j-1}}\Lambda_{j, j - 1}\mathcal{Z}_{j - 1} \, .
\end{equation}
The {\it real time propagator of the ISPI scheme} is introduced as 
\begin{equation}
	\Lambda_{j, j - 1}=F_j\prod_{\s}
\prod_{X = B, D}\det\{\boxx{X}{j,j}{\s}
-\boxx{X}{j,j - 1}{\s}\boxx{X}{j - 1,j - 1}{\s, -1}\boxx{X}{j - 1,j}{\s}
\} \, ,
\end{equation}
with the chosen initial configuration
\begin{equation}
\mathcal{Z}_{1}=F_1\prod_{\s}\prod_{X = B, D}\det
\boxx{X}{1, 1}{\s} \, .
\end{equation}
Furthermore,  we use the definition $\{\tau, \zeta \}_j = (\tau_{j K}^{\mp},
\ldots, \tau_{(j -1) K
+ 1}^{\mp}, \zeta_{j K}^{\mp}, \ldots, \zeta_{(j -1) K + 1}^{\mp})$
  as the tuple of those impurity- and HS-spins that lie in the $j$-th
path segment of length $K$, see Fig.~\ref{fig:SpinPathSegmentation}. The
propagator $\Lambda_{j, j - 1}$ 
(and matrix blocks) depends on all HS- and impurity spins in segments $j - 1$
and $j$. The prefactor 
\begin{equation} \label{eqn:fj}
F_j=2^{-2 K}(-1)^{\ell_j}\left(\frac{J \dt}{2}\right)^{m_j}
e^{i \Phi_{\text{imp}}^{(j)}} \, 
\end{equation}
is related to the number and the position of the flip-flops. $m_j$ is the number
of flip-flops in segment $j$, out of which $\ell_{j}$  lie
on the backward branch. The phase is  defined as 
\begin{equation}
\Phi_{\text{imp}}^{(j)}=-\frac{\Delta_{\text{imp}} \dt}{2} \sum_{l =  (j
- 1)K + 1}^{j K} ( \tau_l^+ - \tau_l^- )\, .
\end{equation}
\begin{figure}
\includegraphics{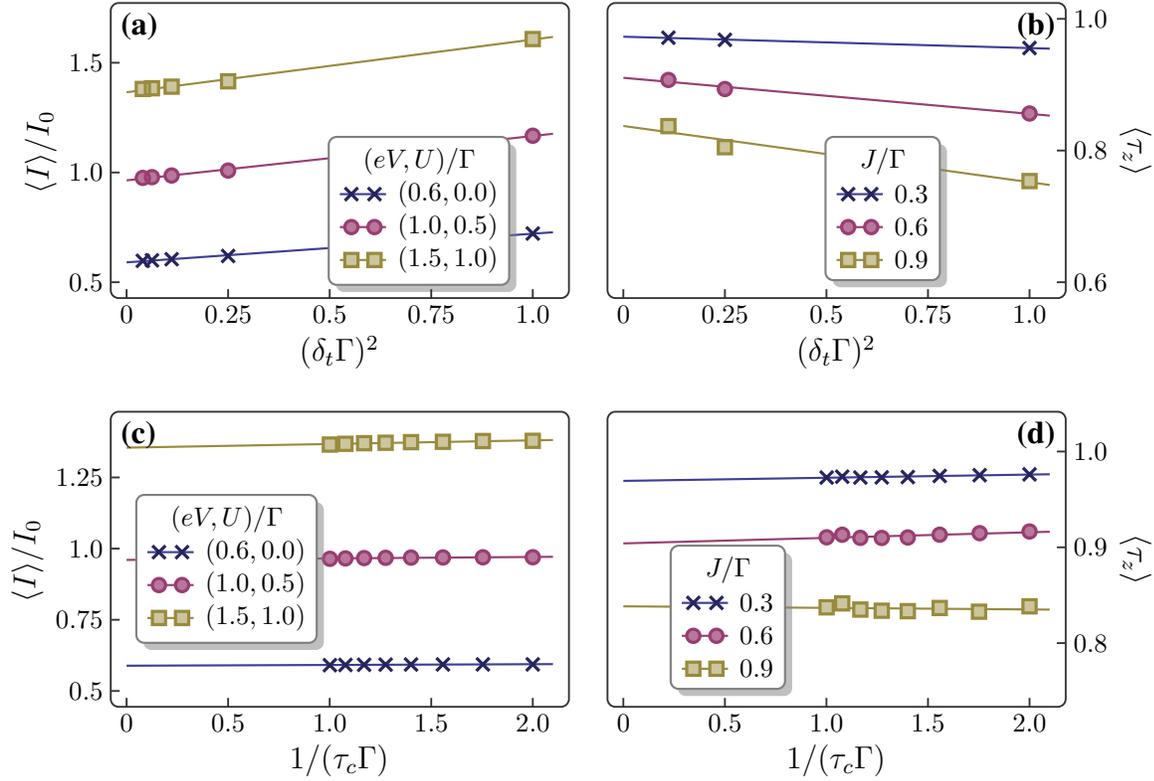}
\caption{
	\label{fig:Extrapolate}
	(a) and (b):
	Trotter extrapolation (solid lines) for $\dt \to 0$  for $\TauC = 1 /\G$,
	$t_{m} \G = 4$, $\b \G = 1$, and $\Phi_D = \Delta = \Delta_{\text{imp}} = 0$.
	(a) Charge current in units of $I_0 = e \G / h$ for $J = 0$, (b) mean impurity
	spin polarisation for $e V = 0.5 \G$ and $U = 0.5 \G$. The impurity spin was
	initially in the spin-up state $\tau_i = 1$. Tiny deviations point to
	negligible unsystematic errors. For all cases in (a), the standard deviations
	are below $1\%$, while in  (b), they rise from around $1\%$ for $J = 0.3 \G$ to
	about $10\%$ for $J = 0.9 \G$.
	(c) and (d):
	Memory extrapolation $\tau_{c}^{-1}\to 0$ for (c) the current and (d) the
	impurity orientation. For all combinations of $e V$ and $U$, the standard
	deviations to the linear fit are below $1\%$. Within the error margin,
	the numerical value of the current for $e V = 0.6 \G$ and $U = 0$ coincides with
	the Landauer-B\"uttiker value $I_{\text{LB}} \approx 0.594 I_0$.
}
\end{figure}

\subsection{Extrapolation procedure} \label{subsec:Extrapolation}
By construction, the numerical value of $\mathcal{Z}[\eta]$ contains two
systematic errors, namely the finite discretisation time step $\dt$ and the
finite
correlation time $\TauC = (K - 1)\dt$. In the limit $\dt \to 0$ and $K\to
\infty$, however, the iterative procedure yields an exact result. The major
benefit of this iterative scheme is that the numerical costs scale linearly with
evolution time $t_{m} - t_i$. The systematic errors are eliminated
\cite{weiss:195316} by an extrapolation of the numerical 
results to vanishing Trotter increment $\dt$ and infinite memory time $\TauC\to
\infty$. 

Due to the Trotter time discretisation, all expressions are by construction
exact up to order $\dt^{2}$ terms. For a fixed  $\TauC$ and small enough values
of $\dt$, we extrapolate the numerical values of some observable to $\dt \to 0$
and thus completely eliminate the Trotter error. An example is shown in Fig.\
\ref{fig:Extrapolate} (a) and (b) for the current and the impurity orientation,
respectively. Depending on the observable, Trotter convergence may be achieved
on different scales \cite{Weiss2005}. Note that one source of errors is the
numerical derivative in Eq.\ (\ref{eqn:CancelDivergence}) which results in tiny
imaginary parts of observables, ranging between $10^{-5}$ to $10^{-3}$. For
typical parameters, it is at least one order of magnitude smaller than the
numerical error from the linear extrapolation.

\begin{figure}
\includegraphics{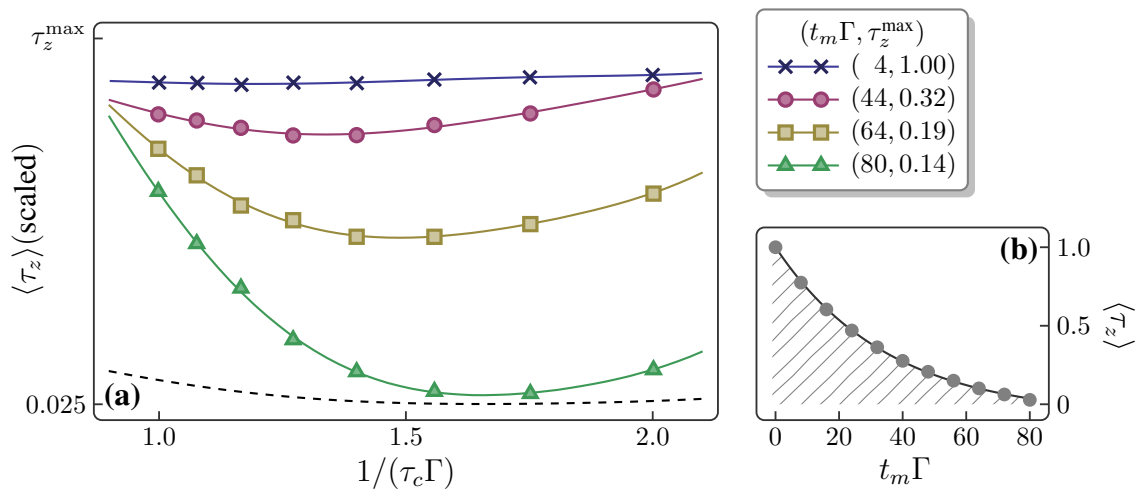}
\caption{\label{fig:LeastDependence}
(a) Mean impurity polarisation vs. $\tau_{c}^{-1}$ 
for different measurement times $t_{m}$ for $\Phi_D = -\G / 2$, $\Delta = \G /
2$, $J
= -\G / 2$, $\b \G = 5$ and $\Delta_{\text{imp}} = U = 0$.  Notice that the
vertical scale is different for each $t_{m}$ to stress the relative variations.
For all plots, the lower bound of the scale is set to $\tau_{z}^{\text{min}} =
0.025$, while the upper bound $\tau_{z}^{\text{max}}$ is as indicated 
(solid lines are guides to the eye). The dashed line illustrates the 
results for $t_{m} \G = 80$ (green triangles) when scaled as the data for $t_{m}
\G =
4$ (blue crosses). The optimal values of $\expect{\tau_z}$ evaluated at the
local minimum are
shown in 
(b) as function of $t_{m}$. A fit to an exponential 
with a relative standard deviation of $5\cdot10^{-3}$ is represented by the
solid line.}
\end{figure}

For the memory extrapolation $\TauC \to \infty$, we do not have a
strict mathematical argument at hand, in contrast to the Trotter extrapolation. 
Whenever results are convergent, however, we find empirically two typical
behaviours:   
(i) either the numerical results for $\expect{O} (\TauC)$ depend linearly 
 on $1/\TauC$ with small deviations, (ii) or their dependence on
$1/\TauC$ is reasonably smooth and exhibits a local extremum. The latter
case is consistent with the principle of least dependence, see
Refs.~\cite{weiss:195316, huetzen2012,  Thorwart00, Eckel06} 
for a verification when compared to analytical results. An example of the linear
scaling of the numerical results with $\tau_{c}^{-1}$
is illustrated in Fig.\ \ref{fig:Extrapolate} (c) and (d).
When $\expect{O}(\tau_{c}^{-1})$ shows a weak dependence on
$\tau_{c}^{-1}$ in a certain corner of parameter space, we still try to apply
criterion (ii). Such a behaviour results from a trade-off between accuracy and
computational costs. A minimal Trotter error requires minimal $\dt$, while, at
the same time, a maximally large correlation time $\TauC$ is desirable.
Naturally,
these requirements are limited by the exponentially increasing numerical costs.
This is illustrated in Fig.~\ref{fig:LeastDependence}.

\subsection{Restricting the number of flip-flops in the memory window}

In order to reduce the exponentially growing number of contributing paths ($\sim
4^K$) without affecting the accuracy, we may exploit that $F_j\sim (J \dt /
2)^{m_j}$ in Eq.\ \eqref{eqn:fj}. The number $m_j$ of flip-flops  in  path
segment $j$ is $0 \le m_{j} \le 2 (K - 1)$.
We observe that the smaller the  weight of each segment is, the more
flip-flops it contains.
On the other hand, the number of path segments $\{\tau\}_{j}$ with $m_{j}$
flip-flops (given by $4 C^{2 (K -1)}_{m_j}$ with $C^n_{k} = n! / [k! (n - k)!]$)
grows as long as $0\le m_{j}\le K - 1$, {\em but decreases again} when $K \le
m_{j} \le 2 (K - 1)$. As a consequence, for any observable there exists a
maximal $m_j^{max}$ such that contributions from paths with $m_j>m_j^{max} \le
2(K-1)$ could safely be disregarded in the numerical iteration. Of course
$m_j^{max}$ is chosen depending on the model parameters and the observable under
investigation.

\begin{figure}
\begin{center}
\includegraphics{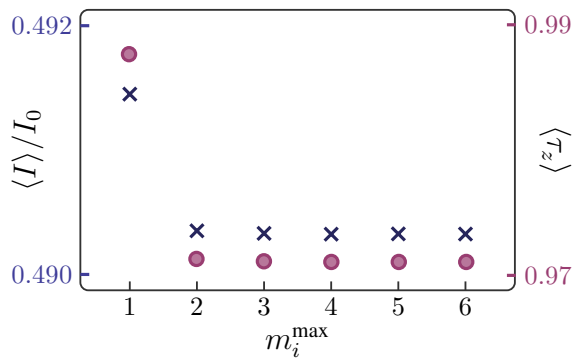}
\end{center}
\caption{\label{fig:ReduceImpurityPaths}
Current (crosses) and impurity spin polarisation (circles) for  $t_{m}
\G = 4$, $\b \G = 1$, $\Phi_D = \Delta = \Delta_{\text{imp}} = 0$, $\dt \G =
1/3$, $K = 4$, and increasing values of the flip-flop number
$m_{j}^{\text{max}}$. Convergence is obtained already for $m_{j}^{\text{max}} =
2$ (notice that the maximal flip-flop number is $2(K-1)=6$). }
\end{figure}

Rapidly decreasing weights of the paths
may not be (over-)compensated by increasing weights for $0 \le m_{j} \le K
- 1$, since each contribution is small and the number
of paths decreases again for larger $m_{j}\ge K$.
The behaviour of the impurity weights is illustrated as follows. Consider 
the case when $m_j$ is close to the maximum $2 (K - 1)$. Both path
classes with $m_{j} = 0$ and $m_{j} = 2 (K - 1)$ contain the same number of
elements (four), while each path contribution in the second class
is weighted by $(J \dt / 2)^{2 (K - 1)}$. , For typical values of $K =
4$, $\dt \G = 1/2$, and $J = \G$, the weight is $\sim 2.5 \times 10^{-4}$. This
also holds for all $K \le m_{j} \le 2 (K - 1)$. Since $m_{j}^{\text{max}}$ is 
unknown {\em a priori\/}, we include it into our code as an additional
parameter.
Then, we perform a numerical estimate 
by a spot sample of the parameter space. It turns out that for many
cases, it is sufficient already to choose $m_{j}^{\text{max}} = 2$.
Fig.~\ref{fig:ReduceImpurityPaths} 
shows an example where both $\expect{I}$
and $\expect{\tau_z}$ converge quickly for increasing $m_{j}^{\text{max}}$. 
This drastically reduces the CPU running times, e.g., for parameters chosen as
in 
Fig.~\ref{fig:ReduceImpurityPaths}, from more than one month to typically three
to five days.

\section{Charge Current and Impurity Dynamics} \label{sec:results}

The ISPI scheme has originally been developed for the Coulomb-interacting
single-lev\-el quantum dot (SLQD). 
By reproducing established analytical and experimentally confirmed results, the
general validity of the 
approach has been shown in Refs.~\cite{weiss:195316, huetzen2012}.
Here, we focus on novel transport features caused by the magnetic impurity  and
its interaction with dot electrons.
\begin{figure}
\begin{center}
	\includegraphics{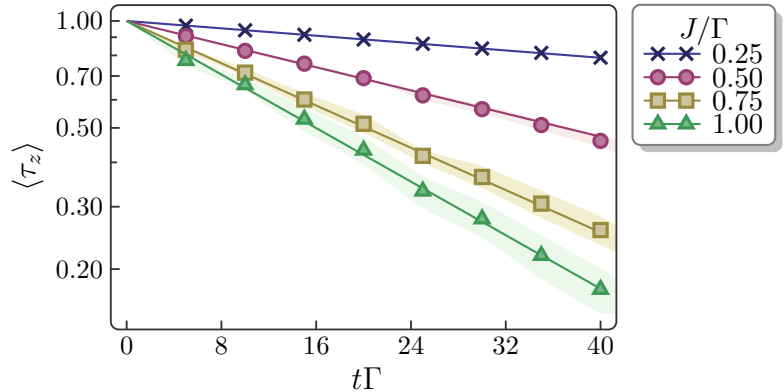}
\end{center}
\vskip-\lastskip
\caption
{
	\label{fig:TimeDecayOfImpSpin}
	The expectation value $\expect{\tau_z}$ (log scale) of the impurity
orientation as a function of time for four different strengths $J$ of an
anti-ferromagnetic electron-impurity interaction. The initial preparation of the
system at $t=-\infty$ is spin-up [$\tau_z (0) \equiv \tau_i = 1$] and we set
$\Phi_D = \Delta = \Delta_{\text{imp}} = U = 0$, $\b \G = 1$, and $e V = 0.6
\G$. The calculated impurity orientation (plot marks) is fitted with good
accuracy (errors indicated by shaded areas) to exponentially decaying functions
(solid lines).	}
\end{figure}
We emphasize that novel dynamical and transport features are mediated by the
transverse or flip-flop interaction $\op{H}_{\text{int}}^{\perp}$,  given by
Eq.~\eqref{eqn:model:H:int}. Without the possibility of flip-flops the
orientation of the impurity spin and its quantum state could not change and not
participate in the dynamics. 
The remaining longitudinal part of the interaction
$\op{H}_{\text{int}}^{\parallel}$ causes a renormalisation of rates and energies
which adds to the effect stemming from a magnetic field. 
Necessarily, flip-flop processes are
involved from the beginning to investigate the non-trivial impurity dynamics 
by considering the time dependence of the impurity orientation
$\expect{\tau_z}$. 

In all presented results below,  the impurity is initially polarized and then
the coupling to the leads is switched on. 
We observe that the time-decay of the polarisation is well described by a single
exponential with a constant decay rate. 
In order to single out the relevant physical processes, we compare our numerical
ISPI results to a 
diagrammatic perturbation theory in the weak- to intermediate exchange
interaction regime. 

We show that for the appropriate parameter regime, 
the exact numerical results are in accordance with the perturbative result and,
by this, we obtain a first intuitive explanation of the impurity dynamics. A
next step is the transfer of its plausibility to the ISPI results. However,
interaction-induced deviations from the perturbative theory are large enough to
clearly illustrate the need for a non-perturbative theoretical description,
provided by the ISPI results.

\subsection{Real-time decay of the impurity polarisation} \label{sec:RateEqImpSpin}

In Fig.~\ref{fig:TimeDecayOfImpSpin} we present the time evolution of the impurity polarisation
 $\expect{\tau_z}$ for different values of the exchange interaction $J$. 
The remaining model parameters are chosen as $\Phi_D = \Delta =
\Delta_{\text{imp}} = U = 0$, and $\b \G = 1$ as well as $e V = 0.6 \G$. 
As a function of time, the impurity polarisation shows a decaying behaviour,
well described by an exponential relaxation for intermediate to long propagation
times.
A faster decay of the polarisation is observed as the impurity interacts
stronger with the electron spins. In passing, we note that a rate equation
ansatz with constant rate $\tau_{\text{R}}^{-1}$, where $\tau_{\text{R}}$ is the
relaxation time, results in an exponential decay as well, i.e., 
$
\expect{\tau_z} (t) \propto e^{-(t - t_i)\tau_{\text{R}}^{-1}}.
$
The parameters are chosen to yield an isotropic (symmetric with respect to
[relative] spin orientations) model system. In this case the anti-ferromagnetic
interaction favours anti-parallel orientation of electron- and impurity spin.
Over long propagation times, the coupling to the unpolarised leads then destroys
any polarisation of the impurity. It is therefore reasonable to assume, that the
rates for up- and down flips are equal. In the chosen parameter range, the
polarisation of the impurity in contact with the leads is well described by a
Markovian dynamics, i.e., solely by the time dependent probabilities $\P_{\tau}
(t)$ of finding the impurity in state $\ket{\tau}$ at time $t$. Apparently, this
simple theoretical prediction agrees well with the numerical results, see
Fig.~\ref{fig:TimeDecayOfImpSpin}.
While the impurity interaction energy is comparable to the tunnel coupling and
considerably affects the transport behaviour as we show below (see
Fig.~\ref{fig:JSweepCurrent}), the rather high temperature and bias voltage
nevertheless reduce the relevance of coherent dynamics due to on-dot
interactions to a secondary role.

\begin{figure}

	\begin{center}
		\includegraphics{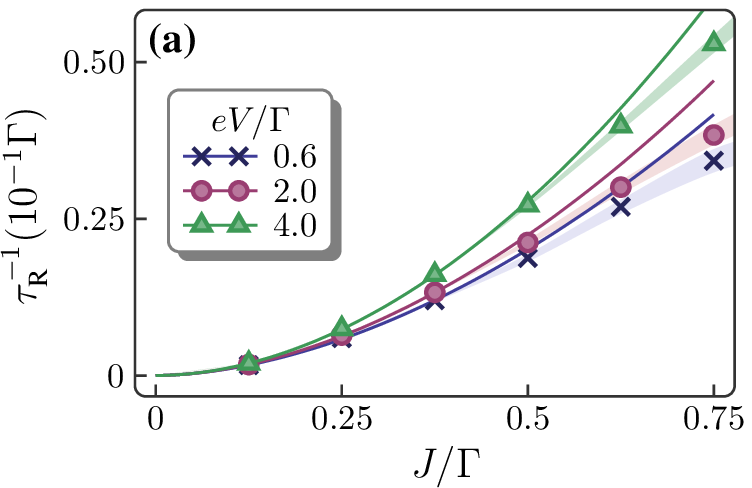}
		\hfill
		\includegraphics{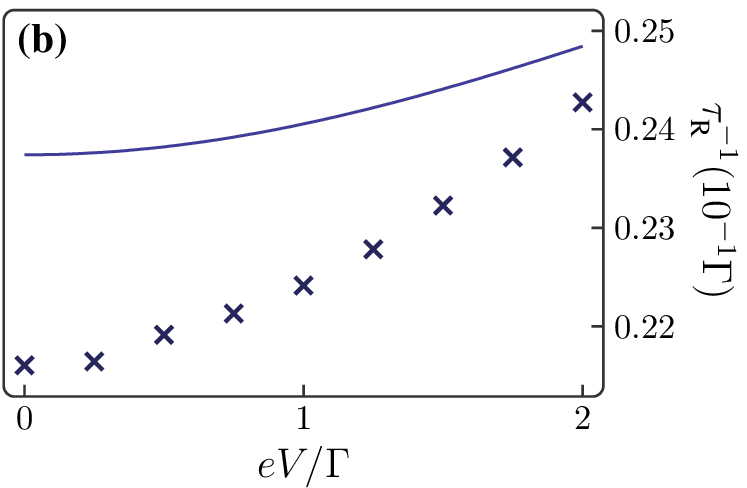}
	\end{center}
	\vskip-\lastskip
	\caption
	{
		\label{fig:SystematicJSweepRelax}
		(a) Impurity relaxation rate $\tau_{\text{R}}^{-1}$ versus
interaction strength $J$, for three different values of $V$. The parameters
correspond to those from Fig. \ref{fig:TimeDecayOfImpSpin}. The solid lines are
fits to polynomial functions $a J^b$ and all resulting values of $b$ are close
to $2$ ($ 1.8\leq b \leq 1.9$). The polarisation decays faster with increasing
$J$. (b) Comparison of the numerically exact [ISPI, crosses] and the sequential 
relaxation rate [Eq.\ \eqref{eqn:InverseRelaxationRate}, blue solid line] versus
bias voltage for $J = \G / 2$ and $\b \G = 1$ (here and in what follows,
parameters not explicitly given are set to zero). Quantitatively, the sequential
and ISPI relaxation times differ noticeably in the crossover
regime.
	}

\end{figure}

 We then turn to the investigation of the inverse relaxation time
$\tau_{\text{R}}^{-1}$. In Fig.~\ref{fig:SystematicJSweepRelax}(a), we present
results for varying $J$ and $U = 0$, and three different bias voltages. These
show a nearly quadratic behaviour  growing from zero (no relaxation) in the sense
that for a fit of the results for $0 \le J \le \G/2$ to a polynomial function $a
J^b$ the exponent $b$ lies between $\sim 1.8$ and $\sim 1.9$.
An \emph{exact} quadratic dependence of $\tau_{\text{R}}^{-1}$ on $J$ is
obtained only in cases where the dynamics is \emph{strongly} dominated by
sequential (incoherent) flip-flop processes. This is only realized when $J \ll
\G$. The corresponding rates may be obtained based on the 
real-time diagrammatic technique developed by Schoeller and Sch\"on
\cite{PhysRevB.50.18436} yielding
\begin{equation}\label{eqn:InverseRelaxationRate}
	\tau_{\text{R}}^{-1}
		\approx
			\frac{J^2 \G^2}{16 \pi}\!\!\!
			\sum_{\alpha, \alpha' = \pm}
			\int_{-\infty}^{\infty} \hspace*{-4mm}\dd \w\;
				\frac
				{
					[f_{\text{L}}^+ (\w) + f_{\text{R}}^+ (\w) ]
					[f_{\text{L}}^- (\w) + f_{\text{R}}^- (\w) ]
				}{
					[(\w - \w_{\uparrow} + \alpha J )^2 + \G^2]
					[(\w - \w_{\downarrow} + \alpha' J)^2 + \G^2]
				}.
\end{equation}
Details of the derivation 
can be found in Ref.\ \cite{Thesis_Becker11}. It reveals the physical structure
and allows for the intuitive interpretation of the processes contributing to
sequential flip-flops. In the numerator of the integrand, we have the sum of all
four possible ways to multiply one of the lead's occupations ($f^+$) with
another or the same lead's probability to find an empty state ($f^-$) at some
energy. Each of these four combinations is then multiplied by the Lorentzian
spectral density for the two different spin states each shifted by $\pm J$
(longitudinal interaction energy). This suggests the following interpretation: A
sequential flip-flop process consists of three elementary  components: the
actual flip-flop and two tunnelling processes of single electrons with opposite
spin (not necessarily in that order). Since they evolve coherently, these
components form an effective spin-flip process $\ket{\chi, \tau} \to \ket{\chi,
-\tau}$, where $\chi \in \{0, \s, \text{d}\}$ and the underlying flip-flop
nature is masked by the tunnelling electrons.  For a particular choice of
$\alpha$ and $\alpha'$ in Eq.~\eqref{eqn:InverseRelaxationRate}, we assign
certain effective flip processes.

Fig.~\ref{fig:SystematicJSweepRelax}(b) shows, how the ISPI result (blue
crosses) compares to the sequential relaxation time (blue solid line). Although
the latter is of the correct order of magnitude, it is systematically larger
than the exact value by $\gtrsim 10\%$.
Since $J$ is not a small parameter of the system, we can presume that the
deviations are mostly coming from coherent higher-order flip-flop processes,
which are neglected in Eq.~\eqref{eqn:InverseRelaxationRate}. Another source of
those deviations may be that free Green's functions are used for the derivation
of the rates in Eq.~\eqref{eqn:InverseRelaxationRate}. In their qualitative
features, however, both results agree. From their finite value at zero bias
voltage they grow monotonically. While for small voltages the relaxation shows a
nearly quadratic functional form (power-law), for larger bias voltages  $e V
\gtrsim 1.25 \G$, it exhibits a linear behaviour. 

\subsection{Impact of the impurity interaction on the current}
\label{subsec:JonCurrent}

\begin{figure}

	\begin{center}
		\includegraphics{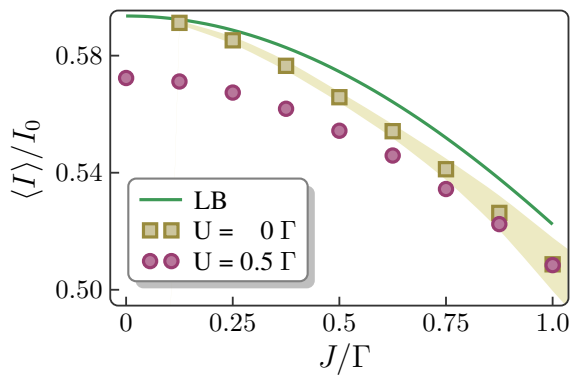}
	\end{center}
	\vskip-\lastskip
	\caption
	{
		\label{fig:JSweepCurrent}
		Stationary charge current $\expect{I}$ in units of $I_0 = e \G /
h$ at $e V = 0.6 \G$ against $J$ for two values of $U$. For increasing
interaction, both the ($U = 0$) Landauer-B\"uttiker theory and the numerics
predict a decrease of the current in the range $0 \le J \le \G$. The similar
characteristics of the LB curve and the ISPI data points suggest that the
current is mainly affected by the longitudinal part of the electron-impurity
interaction. This is probably due to the relatively high temperature and,
consequently, a short coherence time, which strongly limits the influence of
coherent dynamics. While the LB theory and ISPI agree for $J = \G / 8$, growing
differences for increasing $J$ show the effect of flip-flop processes. The
current for small finite Coulomb interaction (no error bars given), though
consistently smaller than the LB values and also decreasing with $J$, drops
slower than for vanishing $U$.
	}

\end{figure}

 As opposed to the slow impurity dynamics, measured in terms of $\G^{-1}$, the
current is relaxing fast into the stationary state. This behaviour is caused by
the strong coupling between the leads and the dot. For the parameters considered
here, the \emph{upper} limit for reaching stationarity is about $t^{\text{ST}}
\lesssim \G^{-1}$. Therefore, we consider the stationary current.
Fig.~\ref{fig:JSweepCurrent} depicts the current as a function of $J$ with $V =
0.6 \G$ and $\b \G = 1$ both for vanishing Coulomb interaction and for $U = \G /
2$. The current decreases with stronger impurity interaction. To distinguish the
influence of the longitudinal (single-particle) and transversal
(spin-scattering) part of the interaction, we compare the ISPI results with the
Landauer-B\"uttiker (LB) current $\expect{I}_{\text{LB}}$ (see
\cite{PhysRevLett.68.2512}),  which can be written here as 
\begin{equation}\label{eqn:LBCurrent}
		\expect{I}_{\text{LB}}
	=	\frac{e \G^2}{2 h}
		\sum_{\s,\alpha=\pm}
		\int_{-\infty}^{\infty} \!\!\!\!\dd \w\,
		\frac
		{
				f^+_{\text{L}} (\w)
			-	f^+_{\text{R}} (\w)
		}{
			(\w - \w_{\s} + \alpha J)^2 + \G^2
		}.
\end{equation}
For $J \ne 0$ this is an approximate expression, as it only includes the effect
of the longitudinal impurity interaction, which acts as an effective magnetic
field in the sense of a mean field. Similar to the sequential relaxation rates
of Eq.~\eqref{eqn:InverseRelaxationRate}, the current formula has a simple
physical interpretation. The joint density of dot-electron states is given by a
Breit-Wigner function, whose width equals the tunnel coupling strength and
whose resonance lies at the single-electron energy $\w_{\s} \pm J$. Hence, the
(non-interacting) current is given by the integral over the energy-dependent
difference of the left and right lead's occupation multiplied by the density of
available dot states at that same energy. The difference in occupation of the
lead electronic states is largest around the Fermi level, where it has the
biggest overlap with the density of states for $J = 0$. With increasing $J$, the
density resonances ``move away'' from the Fermi level, where $f^+_{\text{L}}
(\w) - f^+_{\text{R}} (\w)$ decreases and the current drops.  This effect is
explained in terms of the single-particle energy shift due to the longitudinal
component of the impurity interaction only.

Fig.~\ref{fig:JSweepCurrent} shows that the flip-flop term
$\op{H}_{\text{int}}^{\perp}$ has a considerably smaller influence on the charge
current at this rather large temperature (incoherent regime) than the
longitudinal part of the interaction. Despite the qualitatively similar behaviour
of the LB current and the exact data, the flip-flop scattering causes an
additional significant current drop that grows for growing $J$. A finite Coulomb
interaction of $U = \G / 2$ increases the resistivity of the dot and the ISPI
points are consistently lower than the LB values. The decreasing effect of the
longitudinal part of the impurity, however, is partially compensated by a
broadening of the joint density of states due to Coulomb fluctuations.

\begin{figure}

	\begin{center}
		\includegraphics{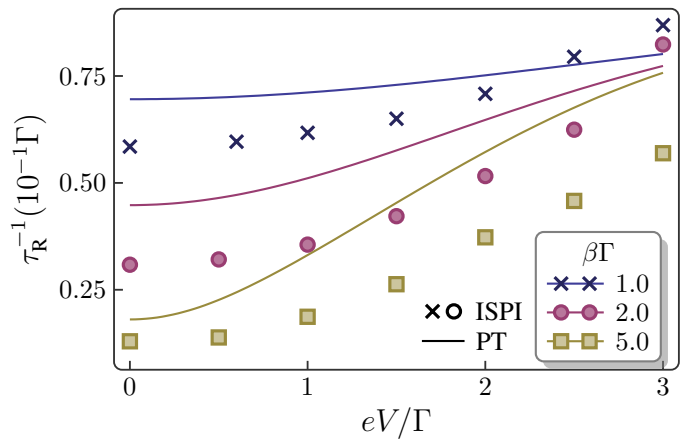}
	\end{center}
	\vskip-\lastskip
	\caption
	{
		\label{fig:VBSweep}
		Relaxation rate versus bias voltage for three different
		temperatures and $U = \G / 2$, $J = \G$. Shown are the ISPI data (symbols) and
		the perturbative (PT) results (solid lines). When compared to the results of
		Fig.~\ref{fig:SystematicJSweepRelax}(b), where $U=0$ and smaller $J$, 
		we see larger relative deviations between the ISPI results and the sequential
		flip-flop approximation. The qualitative differences are also more pronounced.
	}

\end{figure}

\subsection{Finite Impurity Interaction and Coulomb Repulsion}
In the deep quantum regime, where no small parameter exists, ISPI
is certainly applicable and able to describe physical properties
not predictable by perturbative methods. In this section we study, 
how the relaxation rate and the current behave as functions of bias voltage, 
Coulomb interaction and temperature, respectively. 

Figure \ref{fig:VBSweep} presents results in a voltage range $0 \le e V \le 3
\G$, with $J = \G$, $U = \G / 2$ and for temperatures $\b \G = 1$, $2$ and $5$.
The ISPI data of $\tau_{\text{R}}^{-1}$ are indicated by the symbols, while the
solid lines mark the corresponding perturbative rates. The latter exhibit the
same features (power-law growth, followed by a [quasi-]linear behaviour, which
finally saturates) as in Fig.~\ref{fig:SystematicJSweepRelax}(b), which are more
pronounced for lower temperatures. As expected, the ISPI data points deviate
considerably from this lowest-order approximation. Quantum coherent effects are
increasingly relevant since, all energies are of the same order. Both, the
degree of quantitative differences and the deviations in the qualitative
behaviour increase with lower temperatures.

\begin{figure}

	\begin{center}
		\includegraphics{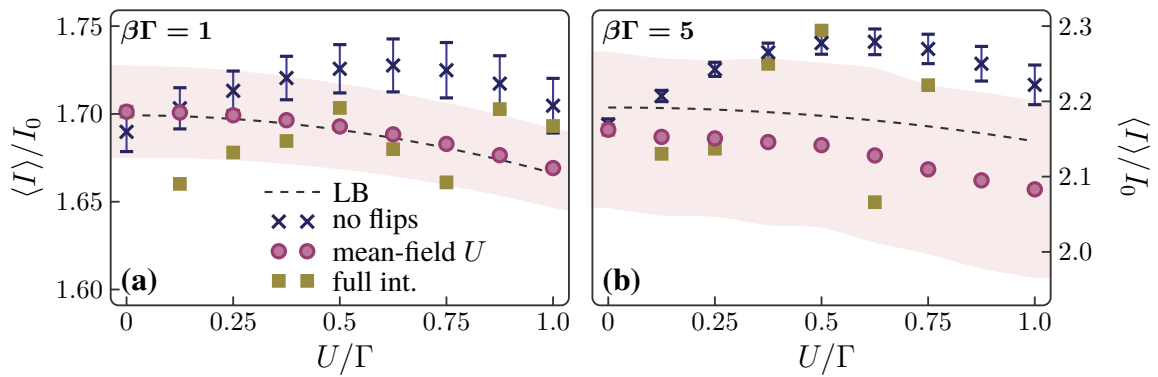}
	\end{center}
	\vskip-\lastskip
	\caption
	{
		\label{fig:USweepCurrent}
		Comparison of (i) the LB current (dashed lines), (ii) the
		Coulomb interacting current without flip-flop scattering (``no flips''), (iii)
		the current without Coulomb scattering but full impurity interaction
		(``mean-field $U$'', see text), and (iv) the fully interacting
		current (``full int.'') in their dependence on the Coulomb interaction $U$ for
		(a) $\b \G = 1$ and (b) $\b \G = 5$. The other (non-zero) parameters are $J =
		\G$ and $V = 2 \G$. The red-shaded areas indicate the error margin for the case
		``mean-field $U$''. 
	}
\end{figure}

In Fig.~\ref{fig:USweepCurrent}, for each of two different temperatures four
different current curves are shown---one for each possibility to either have (i)
only mean field dynamics (LB), (ii) the full Coulomb interaction without
flip-flop processes (``no flips''), (iii) flip-flop dynamics without Coulomb
fluctuations (``mean-field $U$''), or (iv) the fully interacting
dot (``full int.''). For $J = \G$ and $V = 2 \G$, the Coulomb energy is varied
between $0 \le U \le \G$. The situation ``mean-field $U$'' is
implemented by setting $\Phi_D=U / 2$ and the HS-parameter $\lambda=0$ to
illustrate the effect of the ``classical`` or mean-field part of
the Coulomb interaction. This is tantamount to setting the third term in
Eq.~\eqref{eqn:EffectiveDotEnergy} to zero thereby neglecting the 
Coulomb interaction induced fluctuations, which results in
a shift of each single electron energy  by $U/2$. Since
this shift tends to move the transport channels away from the Fermi level, i.e., the
region with the highest density of state in the leads, the current drops. By
its nature, this decrease is equivalent to the one observed for the LB current.%

Only for the ``single-interaction'' currents (``no flips''), we show the error
bars. The reason why no margin of confidence is given for the fully interacting
case, regards the comparability of the error data. Calculating the ``full int.''
current is a very time consuming task and thus, the extrapolation involves
considerably fewer data points. Nevertheless, this does not render these values
unreliable. We still see a compelling linear behaviour of the $1/\TauC$
extrapolation with errors of the order of $1\%$ based on the sample standard
deviation.
	Notice that with about $10\%$, the relative error of the current
values is rather small, the small variations of the
``full int.'' data are
solely due to the extrapolation errors and have no physical meaning.%

For both temperatures, both the LB current and the current without Coulomb
scattering show only a weak dependence on $U$ due to the single-particle energy
shift. The current with full Coulomb interaction but fixed impurity shows a
local maximum for $U \sim \G / 2$, which is more pronounced for $\b \G = 5$. In
this case, the fixed impurity acts as an effective static magnetic field. The
ISPI values for the fully interacting dot vary strongly over the considered $U$
interval, but are scattered around the ``no flips'' and ``no $U$'' curves.

As long as the Coulomb interaction is small, all current values in both figures
lie close. The difference between the respective current values is
given by the inclusion or exclusion of flip-flop processes only.
Hence the rather good agreement of the $U = 0$ values suggests, that even at this temperature
flip-flop processes alone affect the current only weakly. Nevertheless, for
lower temperatures, the flip-flop processes start to influence the current
more strongly, which results in an increased resistivity. The case of the ``no
flip'' current (fixed impurity) is equivalent to a Coulomb-interacting SLQD in a
magnetic field. Both curves in (a) and (b) show a very similar dependence on
$U$, featuring a local maximum at around $U = \G / 2$. For the lower
temperature, the relative height of the broad current peak is twice as big as
for $\b \G = 1$. This effect is also caused by the broadening of the dot's joint
density of states due to the Coulomb fluctuations.

\section{Conclusions}

We have studied the real-time nonequilibrium dynamics of a single-orbital
magnetic quantum dot including Coulomb interactions. In order to obtain
stationary nonequilibrium states at asymptotic times, the ISPI scheme is
employed and extended to cases, when an additional magnetic degree of freedom is
present. Besides the appearance of a Hubbard-Stratonovich field, which decouples
the Anderson repulsion term in the Hamiltonian, we have to include the impurity
interaction on the same level. This nontrivial task requires an additional
summation over paths of the impurity spin degree of freedom. The resulting
action in the path integral formalism involves the Green's function of the
quantum dot as well as its inverse. Inversion of the Green's matrix enlarges the
numerical effort tremendously. From the technical point of view, appearing
matrices, dealing with the impurity dynamics may violate the necessary block
diagonal structure. However, a unitary transformation helps to build up proper
ingredients for the algorithm. Then, also impurity induced correlations become
tractable and do not violate the exponential decay of quantum many-body
correlations. We have presented how an efficient truncation scheme provides
accurate results for the coupled spin dynamics. Results are given for a quantum
spin-$1/2$ impurity on the dot, whereas the generalization to an impurity with a
larger spin is possible. This would be necessary when, e.g. a Mn system is under
investigation.

For the same kind of exchange coupling, the implementation of a Mn
impurity essentially increases the dimension of the impurity path sum. 
Instead of summing over all step-like paths of a spin-$1/2$, it
involves the sub-class of step-like paths (steps between orientations differing by one)
in the space of the six possible orientations of spin $5/2$. Of course, 
the numerical efforts also increase but still are within reach of the ISPI method.
Further work will be dedicated to this goal.

The exponential drop of time correlations due to the leads' coupling allows for
an efficient truncation of the appearing sums---the main building block of the
ISPI scheme. Its application yields high quality numerical data for the impurity
relaxation time and the tunnelling current as a function of the bias voltage in
the presence of (Coulomb) and electron-impurity interactions. We have performed
necessary checks to compare our findings to established results. In the regime
of small impurity interaction, where sequential flip-flops dominate the impurity
dynamics, we have found good agreement with a classical rate equation. This is a
useful tool to gain insight into the dominating processes in the incoherent
regime. Relaxation is described reasonably well by a rate equation when
lead-induced coherences are absent. 

In the deep quantum regime, however, we find that the ISPI method is the only
tool to obtain both the correct order of magnitude and the qualitative features
of the relaxation rate as it depends on the system parameters $U$ and $J$. The
same holds for the influence of $J$ and $U$ on the current in the deep quantum
regime. Most importantly, the ISPI scheme proves to be useful to cover the full
cross-over regime where no small parameter can be identified and thus any
perturbative approach becomes invalid. 

Furthermore, Kondo physics in such a single spin system under
nonequilibrium conditions is of course an interesting subject to study. It
emerges when the Coulomb interaction is large compared to the tunnel coupling
and the temperature is sufficiently low (also in comparison to the tunnel
coupling). In the present work, the two interaction terms in the Hamiltonian 
and the strong tunnel coupling presently limit the application of ISPI to
intermediate temperatures. Therefore, Kondo features
have not yet been obtained. In that regard, the further development of the
method is still demanding, see also the discussion in Ref. \cite{weiss:195316}.
Nevertheless, due to the suppression of long-time correlations at finite
voltages, the regime of nonlinear transport where the Kondo correlations
compete with the finite bias is accessible and will be treated in future 
work.

We have provided a first glimpse on the interesting new physics that comes into
reach with the ISPI scheme. A generalisation of the model to several localised
magnetic impurities, with electrons mediating a finite magnetisation between
them should be possible. The real-time dynamics and all-electrical control of
such devices could be simulated. The presented scheme is also applicable to 
provide the $x$- and $y$-components of the impurity spin, thus yielding the
complete spin dynamics and the real time dephasing on the Bloch sphere.

\section*{Acknowledgements}
This work has been supported by the DFG SFB 668 and by the DFG SPP 1243. 
We thank Reinhold Egger, Jens Eckel, Roland H\"utzen and Benjamin Baxevanis 
for useful discussions.


\section*{References}

\begin{thebibliography}{10}

\bibitem{furdyna:R29} Furdyna J K 1988 {\it J. Appl. Phys.}, {\bf 64} R29

\bibitem{0268-1242-17-4-310}  Dietl T 2002 {\it Semicond.\ Sc.\ Techn.} {\bf 17}
377

\bibitem{RevModPhys.78.809}  Jungwirth T and Sinova J, {Ma\ifmmode
\check{s}\else \v{s}\fi{}ek} J,  {Ku\ifmmode \check{c}\else \v{c}\fi{}era} J and
MacDonald A H 2006 {\it Rev. Mod. Phys.}, {\bf 78} 809

\bibitem{Beaulac2008}  Beaulac R, Archer P I, Ochsenbein S T and Gamelin D R
2008 {\it Adv. Func. Mat.}, {\bf 18} 3873

\bibitem{mackowski:3337}  Mackowski S, Gurung T, Nguyen T A, Jackson H E, Smith
L M,  Karczewski G and Kossut J, 2004 {\it Appl. Phys. Lett.}, {\bf 84} 3337

\bibitem{PhysRevB.71.035338}  Govorov A O and Kalameitsev A V 2005 {\it Phys.
Rev. B}, {\bf 71} 035338

\bibitem{0295-5075-81-3-37005}  Cheng S J and Hawrylak P 2008 {\it
Europhys. Lett.}, {\bf 81} 37005

\bibitem{PhysRevLett.102.177403}  Reiter D E, Kuhn T and Axt V M 2009 {\it Phys.
Rev. Lett.}, {\bf 102} 177403

\bibitem{Zutic2009}  Zutic I and Petukhov A 2009 {\it Nature Nanotechnology},
{\bf 4} 623

\bibitem{PhysRevB.81.245315}  {Le Gall} C, Kolodka R S, Cao C L, Boukari H,
Mariette H, Fern{\'a}ndez-Rossier J and Besombes L 2010 {\it Phys. Rev. B}, {\bf
81} 245315

\bibitem{Ochsenbein2009}  Ochsenbein S T, Yong Feng, Whitaker K M, Badaeva E, 
Liu W K, Li X and Gamelin D R 2009 {\it Nature Nanotechnology}, {\bf 4} 681

\bibitem{PhysRevLett.97.107401}  L{\'e}ger  Y, Besombes L, Fern{\'a}ndez-Rossier
J, Maingault L and  Mariette H 2006 {\it Phys. Rev. Lett.}, {\bf 97} 107401

\bibitem{PhysRevLett.98.106805}  Fern{\'a}ndez-Rossier J and Ram{\'o}n Aguado
2007 {\it Phys. Rev. Lett.}, {\bf 98} 106805

\bibitem{Hanson2008} Hanson R and Awschalom D D 2008 {\it Nature}, {\bf 453}
1043

\bibitem{Wiebe07}
Marczinowski F, Wiebe J, Tang J-M, Flatt\'{e} M E, Meier F, 
Morgenstern M and Wiesendanger R 2007 {\em Phys. Rev. Lett.}, 
{\bf 99},  157202

\bibitem{Khajetoorians10}
Khajetoorians A A, Chilian B, Wiebe J, Schuwalow S, Lechermann F and
Wiesendanger R 2010 {\em Nature} {\bf 467}, 1084

\bibitem{Khajetoorians11}
Khajetoorians A A, Lounis S, Chilian B, Costa A T, Zhou L, Mills D L,
Wiebe J and Wiesendanger R 2011  {\em Phys. Rev. Lett.}, 
{\bf 106},  037205

\bibitem{Herrmann}
Herrmann C, Solomon G C and Ratner M A 2010 {\em J. Am. Chem. Soc.} {\bf 132}
3682

\bibitem{Berndt}Gopakumar T G, Matino F, Naggert H, Bannwarth A, Tuczek F and 
Berndt R 2012 {\em Angew. Chem. Int. Ed} {\bf 51} 6262

\bibitem{Hanson2007}  Hanson R, Petta J R, Tarucha S and Vandersypen
L M K 2007 {\it Rev. Mod. Phys}, {\bf 79} 1217 

\bibitem{Churchill2009} Churchill H O H, Kuemmeth F, Harlow J W,
Bestwick A J, Rashba E I, Flensberg K, Stwertka C H, Taychtanapat T, Watson S K,
Marcus C M 2009, {\it Phys. Rev. Lett}, {\bf 102} 166802 

\bibitem{kouwenhoven_ganz:1996}  Kouwenhoven L P, Markus C M, McEuen P L,
Tarucha S, Westervelt R M and Wingreen N S 1997 {\it Mesoscopic Electron
Transport} ed L L Sohn {\it et al} (London: Kluwer Academic Publishers)

\bibitem{PhysRevB.77.195416}  Timm C 2008 {\it Phys. Rev. B}, {\bf 77} 195416

\bibitem{PhysRevB.50.18436}  Schoeller H and Sch{\"o}n G 1994 {\it Phys. Rev.
B}, {\bf 50} 18436

\bibitem{PhysRevLett.84.3686}  Schoeller H and K{\"o}nig J 2000 {\it Phys. Rev.
Lett.}, {\bf 84} 3686

\bibitem{PhysRevLett.90.076804}  Rosch A, Paaske J, Kroha J and W{\"o}lfle P
2003 {\it Phys. Rev. Lett.}, {\bf 90} 076804

\bibitem{PhysRevB.69.155330}  Paaske J, Rosch A and W{\"o}lfle P 2004 {\it Phys.
Rev. B}, {\bf 69} 155330

\bibitem{PhysRevLett.95.056602} Kehrein S 2005 {\it Phys. Rev. Lett.}, {\bf 95}
056602

\bibitem{PhysRevB.77.125309}  Weyrauch M and Sibold D 2008 {\it Phys. Rev. B},
{\bf 77} 125309

\bibitem{10.1140Schoeller}  Schoeller H 2009 {\it Eur. Phys. J. -
Special Topics}, {\bf 168} 179 

\bibitem{jakobs:150603}  Jakobs S G, Meden V and Schoeller H 2007 {\it Phys.
Rev. Lett.}, {\bf 99} 150603

\bibitem{PhysRevB.75.045324}  Gezzi R, Pruschke Th and Meden V 2007 {\it Phys.
Rev. B}, {\bf 75} 045324

\bibitem{0295-5075-90-3-30003}  Karrasch C, Andergassen S, Pletyukhov M,
Schuricht D,  Borda L, Meden V and Schoeller H 2010 {\it Europhys. 
Lett.}, {\bf 90} 30003

\bibitem{PhysRevB.81.195109}  Jakobs S G, Pletyukhov M and Schoeller H 2010 {\it
Phys. Rev. B}, {\bf 81} 195109

\bibitem{PhysRevB.79.235336}  Heidrich-Meisner F, Feiguin A E and Dagotto E 2009 {\it
Phys. Rev. B}, {\bf 79} 235336

\bibitem{1742-5468-2004-04-P04005}  Daley A J, Kollath C, Schollw{\"o}ck U and 
Vidal G 2004 {\it J. Stat. Mech.}, {\bf
2004} P04005

\bibitem{PhysRevLett.93.076401}  White S R and Feiguin A E 2004 {\it Phys. Rev.
Lett.}, {\bf 93} 076401

\bibitem{PhysRevB.70.121302}  Schmitteckert P 2004 {\it Phys. Rev. B}, {\bf 70}
121302

\bibitem{muehlbacher:176403}  M{\"u}hlbacher L and Rabani E 2008 {\it Phys. Rev.
Lett.}, {\bf 100} 176403

\bibitem{PhysRevB.78.235110}  Schmidt T L, Werner P, M{\"u}hlbacher L and Komnik
A 2008 {\it Phys. Rev. B}, {\bf 78} 235110

\bibitem{werner:035320}  Werner P, Oka T and Millis A J 2009 {\it Phys. Rev. B},
{\bf 79} 035320

\bibitem{PhysRevB.81.035108}  Werner P, Oka T, Eckstein M and Millis A J 2010
{\it Phys. Rev. B}, {\bf 81} 035108

\bibitem{schiro:153302}  Schir{\'o} M and Fabrizio M 2009 {\it Phys. Rev. B},
{\bf 79} 153302

\bibitem{LichtensteinRMP}Gull E, Millis A J, Lichtenstein A I, Rubtsov A N,
 Troyer M and Werner P 2011 {\it Rev. Mod. Phys.}, {\bf 83} 349 

\bibitem{PhysRevLett.99.236808}  Han J E and Heary R J 2007 {\it Phys. Rev.
Lett.}, {\bf 99} 236808

\bibitem{EckelNJP2010}Eckel J, Heidrich-Meisner F, Jakobs S G, Thorwart M,
Pletyukhov M and Egger R 2010 {\it New J. Phys.}, {\bf 12} 043042

\bibitem{GullPRB2011}  Gull E, Reichman D R and Millis A J 2011,
{\it Phys. Rev. B}, {\bf 84} 085134 

\bibitem{weiss:195316}  Weiss S, Eckel J, Thorwart M and Egger R 2008 {\it Phys.
Rev. B}, {\bf 77} 195316, \textit{ibid.} 2009 {\bf 79}, 249901(E)

\bibitem{huetzen2012}  H\"utzen R, Weiss S, Thorwart M and Egger R  2012 {\it
Phys. Rev. B}, {\bf 85} 121408(R)

\bibitem{PhysRevB.82.205323}  Segal D, Millis A J and Reichman D R 2010 {\it
Phys. Rev. B}, {\bf 82} 205323

\bibitem{Keldysh64}  Keldysh L V 1964 {\it Zh. Eksp. Teor. Fiz.}, {\bf 47} 1515

\bibitem{kamenev-2009-58}  Kamenev A and Levchenko A 2009 {\it Adv. Phys.},
{\bf 58} 197

\bibitem{Thesis_Becker11}  Becker D 2011 {\it Interplay of decoherence and
quantum correlations in  time-dependent transport through small quantum dots}
(Universit{\"a}t Hamburg)


\bibitem{Thorwart00} Thorwart M, Reimann P and H\"anggi P 2000 
{\it Phys. Rev. E}, {\bf 62} 5808 

\bibitem{Eckel06}Eckel J, Weiss S and Thorwart M 2006 {\it Eur. Phys. J. B},
{\bf 53} 91 

\bibitem{PhysRevLett.68.2512}  Meir Y and Wingreen N S 1992 {\it Phys. Rev.
Lett.}, {\bf 68} 2512

\bibitem{Weiss2005}  Weiss S and Egger R 2005 {\it Phys. Rev. B}, {\bf 72}
245301


\end{thebibliography}
\end{document}